\documentclass[%
 reprint,
showpacs,
 amsmath,amssymb,
 aps,pra,mathabx,
floatfix,
]{revtex4-1}

\usepackage{graphicx}
\usepackage{dcolumn}
\usepackage{bm}

\usepackage{epstopdf}
\usepackage{braket}
\usepackage{color}
\usepackage[normalem]{ulem}
\usepackage{soul}
\usepackage{cancel}
\usepackage{verbatim}
\usepackage{notoccite}

\begin{document}

\preprint{APS/123-QED}

\title{STIRAP in sodium vapor with picosecond laser pulses}

\author{Jim L. Hicks}
\affiliation{Department of Chemistry and Physics, Northeastern State University, Talequah, Oklahoma 74464, USA.}
\affiliation{Department of Chemistry and Physics, Northeastern State University, Talequah, Oklahoma 74464, USA.}
\author{Chakree Tanjaroon}
\affiliation{Department of Earth Sciences, The University of Hong Kong, Hong Kong}
\author{Susan D. Allen}%
\affiliation{Embry-Riddle Aeronautical University, Daytona Beach, Florida 32114, USA}
\author{Matt Tilley}
\email{mtilley@uw.edu, Corresponding author}
\affiliation{Department of Earth \& Space Sciences, University of Washington, Seattle, Washington 98195, USA.}
\author{Steven Hoke}
\author{J. Bruce Johnson}
\affiliation{%
 Department of Chemistry and Physics, Arkansas State University, State University, Arkansas 72467, USA}%

\date{\today}

\begin{abstract}
Experimental measurements and calculations of STIRAP transfer efficiencies were made on a sodium gas starting from the $3^2{\rm S}_{1/2}$ electronic ground state, passing through the $3^2{\rm P}_{1/2}$ and/or the $3^2{\rm P}_{3/2}$ to the $5^2{\rm S}_{1/2}$ state. The lasers used in the experiments had a pulse width of several picoseconds and were close to the Fourier transform limit. Although the linewidth of the laser was much smaller than the spin orbit splitting between the $3^2{\rm P}_{1/2}$ and $3^2{\rm P}_{3/2}$ states, Experiments and calculations reveal that both 3p states play a role in the transfer efficiency when the lasers are tuned to resonance through the $3^2{\rm P}_{1/2}$ state, revealing evidence for quantum interference between the competing pathways.


\end{abstract}

\maketitle


\section{\label{Introduction}Introduction}

Stimulated Raman adiabatic passage (STIRAP) and its variations are established techniques for efficient manipulation of population among states of atoms or molecules. STIRAP has been implemented with continuous wave (cw)  \cite{Gaubatz1990} and pulsed \cite{Schiemann1993} lasers for a variety of purposes such as: the preparation of molecules in specific rovibrational states for chemical reactions \cite{Dittmann1992}, preparation of molecules for dissociative attachment of electrons \cite{Kuelz1996,Keil1999}, excitation of targeted magnetic sub levels \cite{Martin1996} and superpositions of levels \cite{Heinz2006,Vewinger2007}, selective momentum transfer for isotopic separation \cite{Theuer1998}, Rydberg state excitation \cite{Kaufmann2001,Cubel2005}, excitation between dipole-forbidden states through a continuum of states \cite{Peters2007}, the coherent preparation of atoms to optimize \cite{Song2007} or measure \cite{Kou2009} coherent anti-Stokes Raman scattering (CARS), excitation of individual Ca$^{2+}$ ions and their potential for information storage \cite{Soerensen2006a}, coherent optical transfer of Feshbach molecules to a lower vibrational state \cite{Winkler2007}, photo association of bi-alkali \cite{Ni2008,Danzl2008}, and other molecules \cite{Stellmer2012}, creating a population inversion in dopants of a solid \cite{Klein2007} including tripod STIRAP in a doped Pr$^{3+}{\rm :Y_2SiO_5}$ \cite{Goto2007}, storing information \cite{Wang2008}, and performing logic operations \cite{Beil2011} in doped solids. 

The majority of work employing STIRAP has been done in atomic or molecular beams or ultracold gases. As a departure from the norm, Stuart Rice proposed the possibility of performing STIRAP in fluids where collisional dephasing could occur \cite{Demirplak2002,Demirplak2006,Masuda2015}. Since collisions in fluids occur on short time scales, short-pulse lasers are required. The work presented here establishes that STIRAP can be performed on a gas with laser pulses several picoseconds in length, a step toward exploring the possibility of performing STIRAP in a fluid. In a paper to follow, we will share our results in performing STIRAP on a gas at atmospheric pressure where collisional dephasing occurs.

To date, only Kuhn et al. studied STIRAP on a gas at or above room temperature \cite{Kuhn1998}. In their pioneering work they demonstrated that for a lambda energy-level configuration, if the initial and final states are somewhat close in energy, STIRAP substantially diminishes the reduction in transfer efficiency due to Doppler broadening (partial Doppler compensation) relative to stimulated emission pumping (SEP). In fact they demonstrated an impressive 18\% STIRAP transfer efficiency (15$\times$ greater than SEP) which was mainly limited by the narrow linewidth of the nanosecond lasers used. 

The work presented here employs picosecond lasers whose shorter pulse width and correspondingly broader bandwidth eliminate the reduction in STIRAP efficiency due to Doppler broadening. In addition, we demonstrate STIRAP on a ladder-ordering of energy levels in atomic sodium vapor for which partial Doppler compensation on a gas at or above room temperature is not possible with nanosecond lasers. 

\section{Background}

\subsection{Sodium}

\begin{figure}[t]
\includegraphics[width=3.5in]{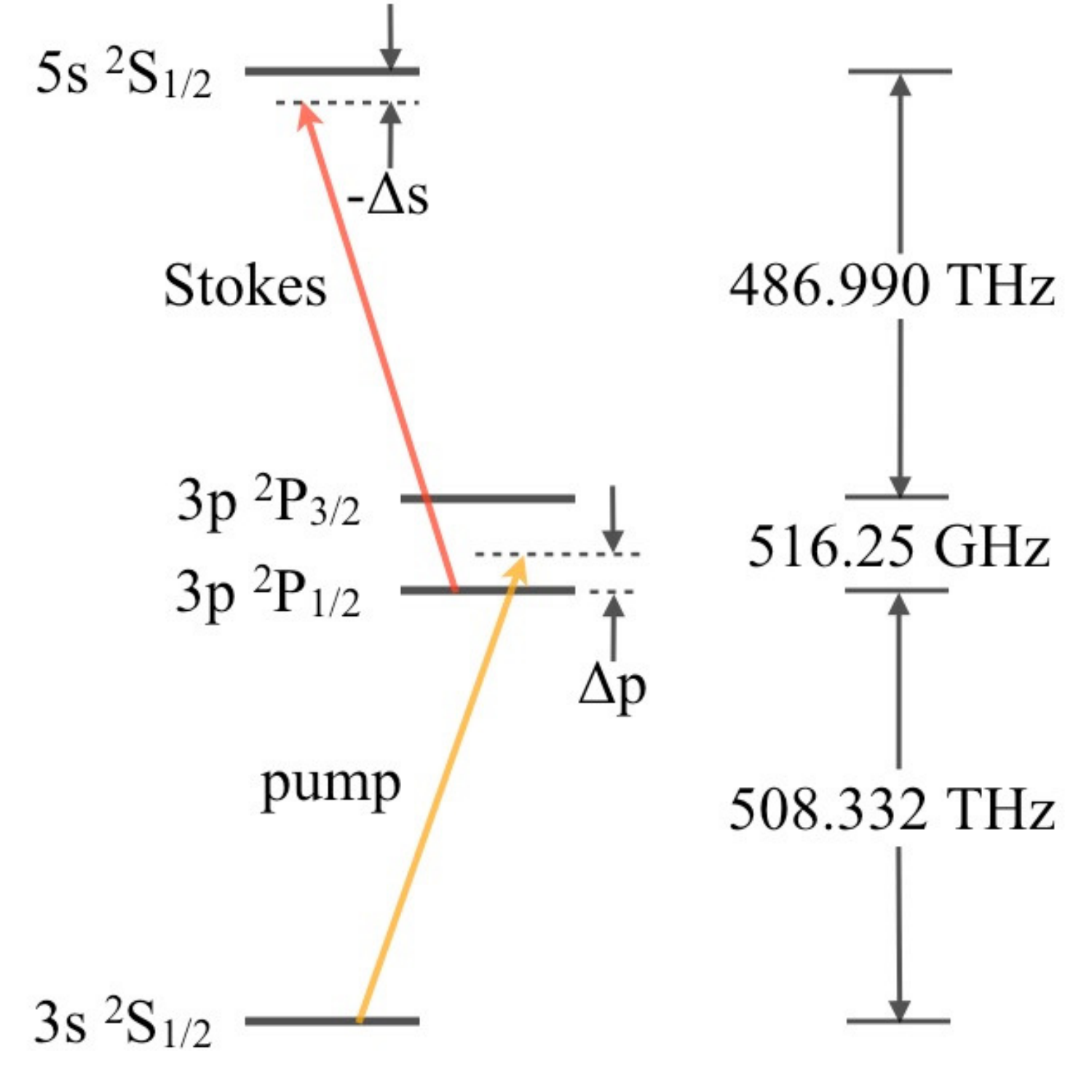}
\caption{\label{fig:Fig_1_Energy_Level_Diagram} (Color online) Sodium energy-levels of the 3s, $3\,^2{\rm P}_{1/2}$, $3\,^2{\rm P}_{3/2}$, and 5s states. Laser detunings for the pump, $\Delta p$, and Stokes, $\Delta s$, are shown here for resonance with the $3\,^2{\rm P}_{1/2}$ state.}
\end{figure}

A diagram of the energy levels used in our excitation of sodium is shown in Fig.~\ref{fig:Fig_1_Energy_Level_Diagram}. The pump laser couples the 3s ($3\,^2{\rm S}_{1/2}$) state to either (or both) of the two 3p levels ($3\,^2{\rm P}_{1/2}$ and $3\,^2{\rm P}_{3/2}$). The Stokes laser couples either (or both) of the 3p levels to the 5s ($5\,^2{\rm S}_{1/2}$) state. In the work presented here, linearly polarized light is used with the polarizations of the pump and Stokes pulses parallel. In this configuration, selection rules dictate that  $\Delta m_J = 0$ and $\Delta J = 0, \pm1$. It follows that the transitions in sodium separate into two independent sets of states (one each for $m_J = \pm 1/2$) that do not interconvert during the excitation process. The Hamiltonian for each of these sets of states is given by 

\begin{equation}
\label{eq:fine}
H_4 = -\frac{\hbar}{2}
\begin{bmatrix}
-2\Delta p & \Omega_{12} & \Omega_{13}  & 0 \\
\Omega_{21} & 0 & 0 & \Omega_{24} \\
\Omega_{31} & 0 & -2\Delta SO & \Omega_{34} \\
0 & \Omega_{42} & \Omega_{43} & 2 \Delta s \\
\end{bmatrix}
\end{equation}
 
 \noindent where $\Delta SO$ is the fine structure splitting, $\Delta p$ and $\Delta s$ are the respective pump and Stokes laser detunings away from resonance, and $\Omega_{i,j}$ is the Rabi frequency between levels $i$ and $j$. The four levels in Eq.~\ref{eq:fine} are the four levels in Fig.~\ref{fig:Fig_1_Energy_Level_Diagram} in ascending order: $\Ket{1}$ is the 3s state, $\Ket{2}$ is the 3p $^2{\rm P}_{1/2}$ state, $\Ket{3}$ is the 3p $^2{\rm P}_{3/2}$ state, and $\Ket{4}$ is the 5s state. The matrix forms of the Hamiltonians for the $m_J = \pm 1/2$ sets of states are identical although the signs of some of the Rabi frequencies differ according the signs of the transition moments. 

The hyperfine levels within the 3$\,^2{\rm S}_{1/2}$, 3$\,^2{\rm P}_{1/2}$, $3\,^2{\rm P}_{3/2}$, and $5\,^2S_{1/2}$ states are treated in Appendix~\ref{app:hyperfine}. All hyperfine splittings are much smaller than the linewidth of the laser (60 GHz). (The largest hyperfine splitting is 1.77 GHz, between the F = 1 and F = 2 states of the $3\,^2{\rm S}_{1/2}$ level. All other splittings are smaller by at least an order of magnitude.) In order to ensure that it is permissible to neglect hyperfine states in the work presented here, calculations are performed with the $4 \times 4$ Hamiltonian above (Eq.~\ref{eq:fine}) and compared with calculations from the two $9 \times 9$ and four $5 \times 5$ Hamiltonians (see Appendix~\ref{app:hyperfine}: Eqs.~\ref{eq:pm2},~\ref{eq:pm1},~\ref{eq:01}, and~\ref{eq:02}) required for hyperfine calculations. In all computations, the calculations that included the hyperfine levels differed from the calculations without by less than 1\%.

\subsection{\label{sec:STIRAP_in_sodium}STIRAP in sodium}

\subsubsection{Resonance with the $3\,^2{\rm P}_{1/2}$ state}

\begin{figure}[t]
\centering
\includegraphics[width=3.5in]{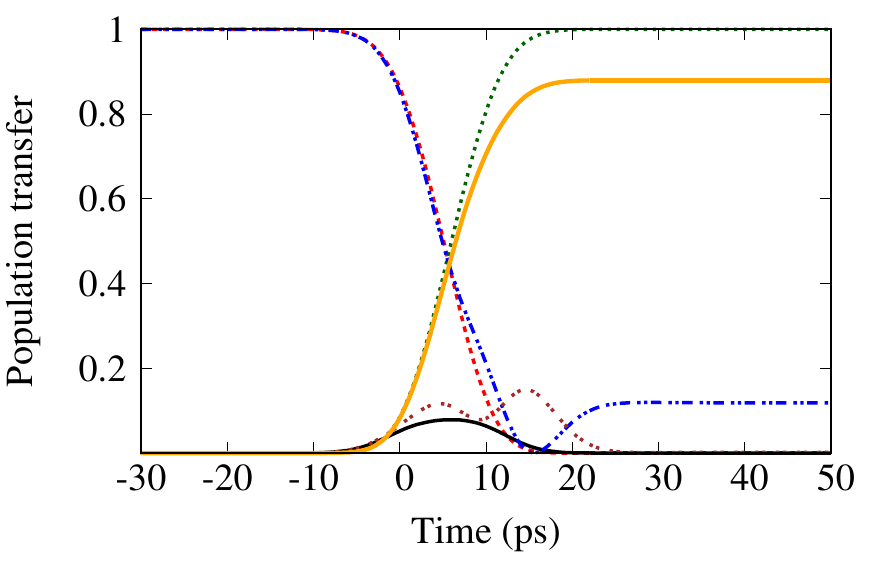}\\
\includegraphics[width=3.5in]{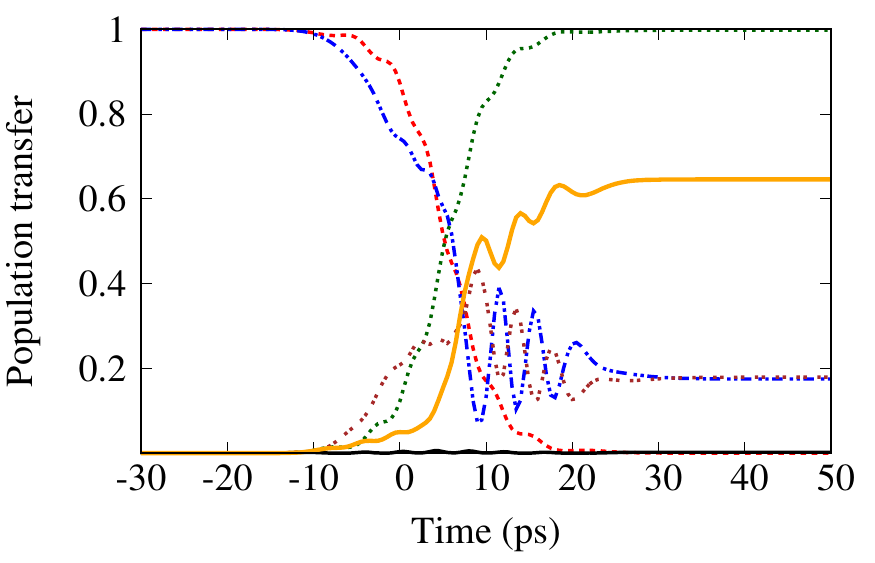}
\caption{\label{fig:Fig_2_STIRAP} (Color online) STIRAP population calculations through one and both 3p states. Relative to the transition moments through  the 3$\,^2{\rm P}_{1/2}$, the integrated Rabi frequencies for the pump and Stokes pulses are 20 rad for the panel on the top and 100 rad for the panel on the bottom. For both panels, the initial ground state population is represented by the dashed red (calculations with $3\,P_{1/2}$ only) and dash-dot-dotted blue (calculations with both $3p$ states) lines, the intermediate state population is represented by the solid black ($3\,P_{1/2}$ only) and dot-dot-spaced brown (both $3p$ states), and the final state population is shown by the dotted green ($3\,P_{1/2}$ only) and solid orange (both $3p$ states) lines.}
\end{figure}

With the lasers tuned to resonance with the $3\,^2{\rm P}_{1/2}$ state, STIRAP calculations were performed that alternately included and neglected the $3\,^2{\rm P}_{3/2}$ state. When neglected, the $4 \times 4$ Hamiltonian given in Eq.~\ref{eq:fine} reduces to a $3 \times 3$ matrix. A comparison of the solutions is shown in Fig.~\ref{fig:Fig_2_STIRAP}. In the top panel, calculations were done with an integrated Rabi frequency of 20 rad. In the bottom panel the integrated Rabi frequency was 100 rad.

In ordinary three-state STIRAP when the pump and Stokes Rabi frequencies are equal, the transfer efficiency to the final state remains near 100\% for integrated Rabi frequencies above a threshold of about 10 rad. Both panels of Fig.~\ref{fig:Fig_2_STIRAP} are consistent with this result when the $3\,^2{\rm P}_{3/2}$ state is excluded from the calculations. By contrast, Fig.~\ref{fig:Fig_2_STIRAP} (top panel) displays that when the $3\,^2{\rm P}_{3/2}$ is included in the calculation the efficiency drops to 88\%. Moreover, Fig.~\ref{fig:Fig_2_STIRAP} (bottom panel) shows that when the integrated Rabi frequency increases from 20 rad to 100 rad, the STIRAP transfer efficiency drops to 65\%. 

These calculations show that the additional intermediate state ($3\,^2{\rm P}_{3/2}$) provides an alternate pathway to the final state, setting up the possibility of quantum interference between the pathways. STIRAP calculations were also performed using Eq.~\ref{eq:fine} in which $\Delta SO$ was varied. For fixed integrated Rabi frequencies of 20 rad, as $\Delta SO$ increased, the STIRAP efficiency increased and asymptotically approached 100\%. This result demonstrates that as $\Delta SO$ becomes large, the $3\,^2{\rm P}_{3/2}$ state decouples from the other states and the laser pulses that excite them, effectively reducing Eq.~\ref{eq:fine} to a $3 \times 3$ Hamiltonian.

It is interesting to note that although the fine-structure splitting of the 3p levels is large (516 GHz) compared with the linewidth of the laser (60 GHz), the Rabi frequencies of the laser pulses were large enough that both 3p levels affected our results. In particular, the lower panel in Fig.~\ref{fig:Fig_2_STIRAP} with a larger integrated Rabi frequency than in the top panel demonstrates that the larger Rabi frequency more effectively couples to both 3p intermediate states, yielding a larger effect (decrease in STIRAP efficiency). 

\begin{figure}[!t]
\includegraphics[width=3.5in]{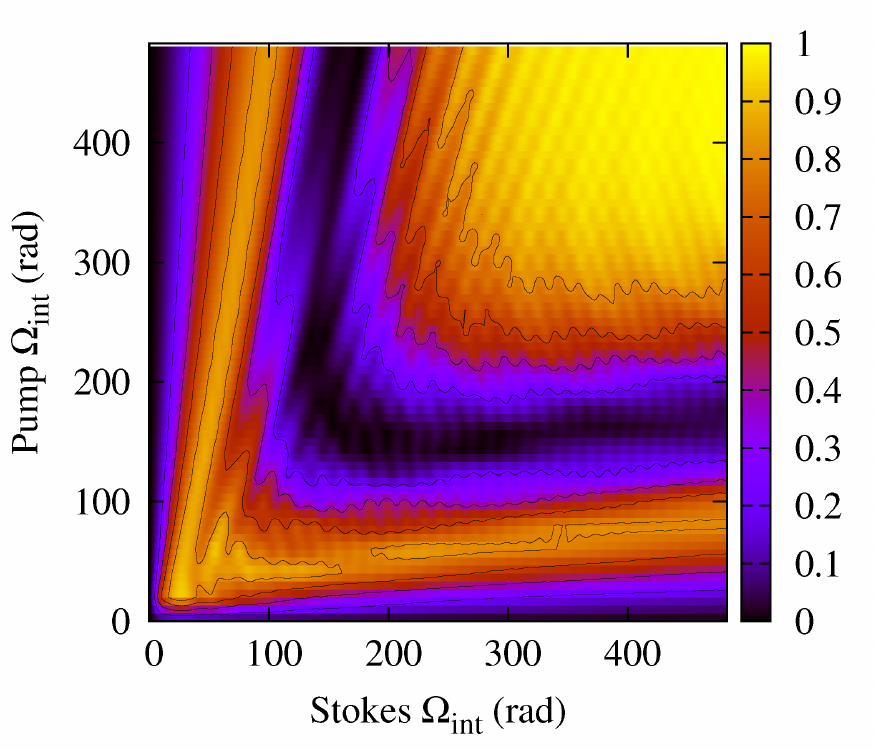}\\
\vspace{-5em}
\includegraphics[width=3.5in]{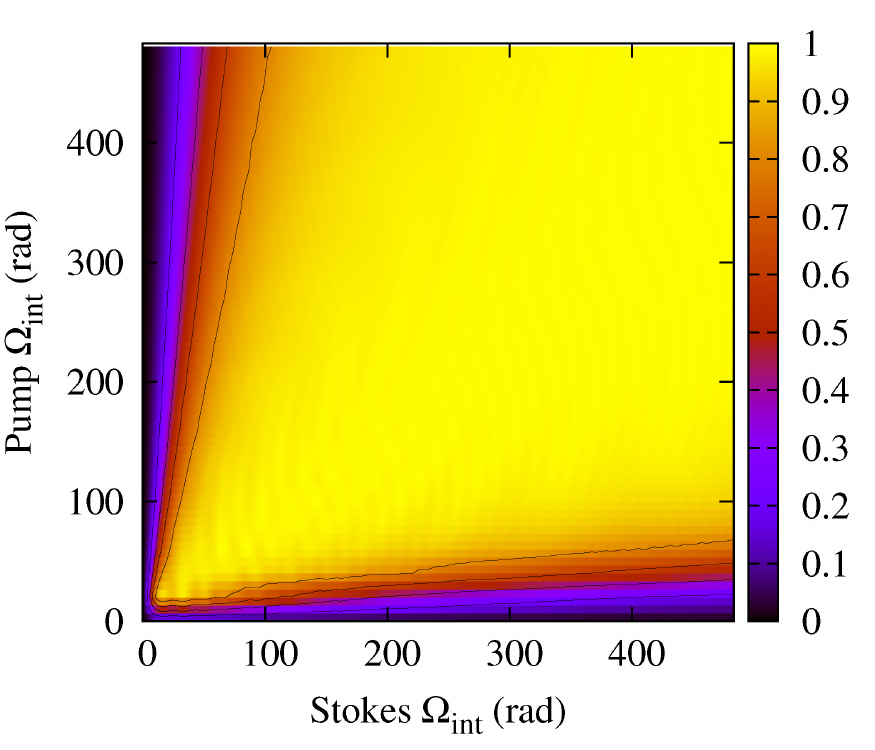}
\caption{\label{fig:Energy_scan_2} (Color online) STIRAP transfer efficiency as a function of the integrated Rabi frequencies of the pump and Stokes pulses. Top panel: The lasers are tuned to resonance through the $3^2{\rm P}_{1/2}$ state. Bottom panel: The same computation is made with the lasers tuned to resonance through the $3^2{\rm P}_{3/2}$ state.}
\end{figure}

A two-dimensional computational study of the transfer efficiency with respect to variations of the pump and Stokes integrated Rabi frequencies $\Omega_{ip}$, $\Omega_{is}$ using Eq.~\ref{eq:fine} was used in order to more fully explore the drop in efficiency with increasing integrated Rabi frequency. The results are shown in Fig.~\ref{fig:Energy_scan_2} (top panel). A significant valley, or dip in the transfer efficiency at integrated Rabi frequencies beyond that of the initial ridge of high transfer efficiency is apparent.

\subsubsection{Resonance with the $3\,^2{\rm P}_{3/2}$ state}

With the lasers tuned to resonance with the $3\,^2{\rm P}_{3/2}$ state, calculations were performed that alternately included and neglected the $3\,^2{\rm P}_{1/2}$ state. The calculated results through both 3p states are shown in Fig.~\ref{fig:Energy_scan_2} (bottom panel). The transfer efficiency remains near unity once the plateau is reached. In this case there is no dip in the transfer efficiency as was evident in Fig.~\ref{fig:Energy_scan_2} (top panel). Additional effects of the interfering pathways will be evident in the results shown later.

\section{Experimental Setup}
\label{sec:exp_setup}

\begin{figure}[t]
\includegraphics[width=4.0in]{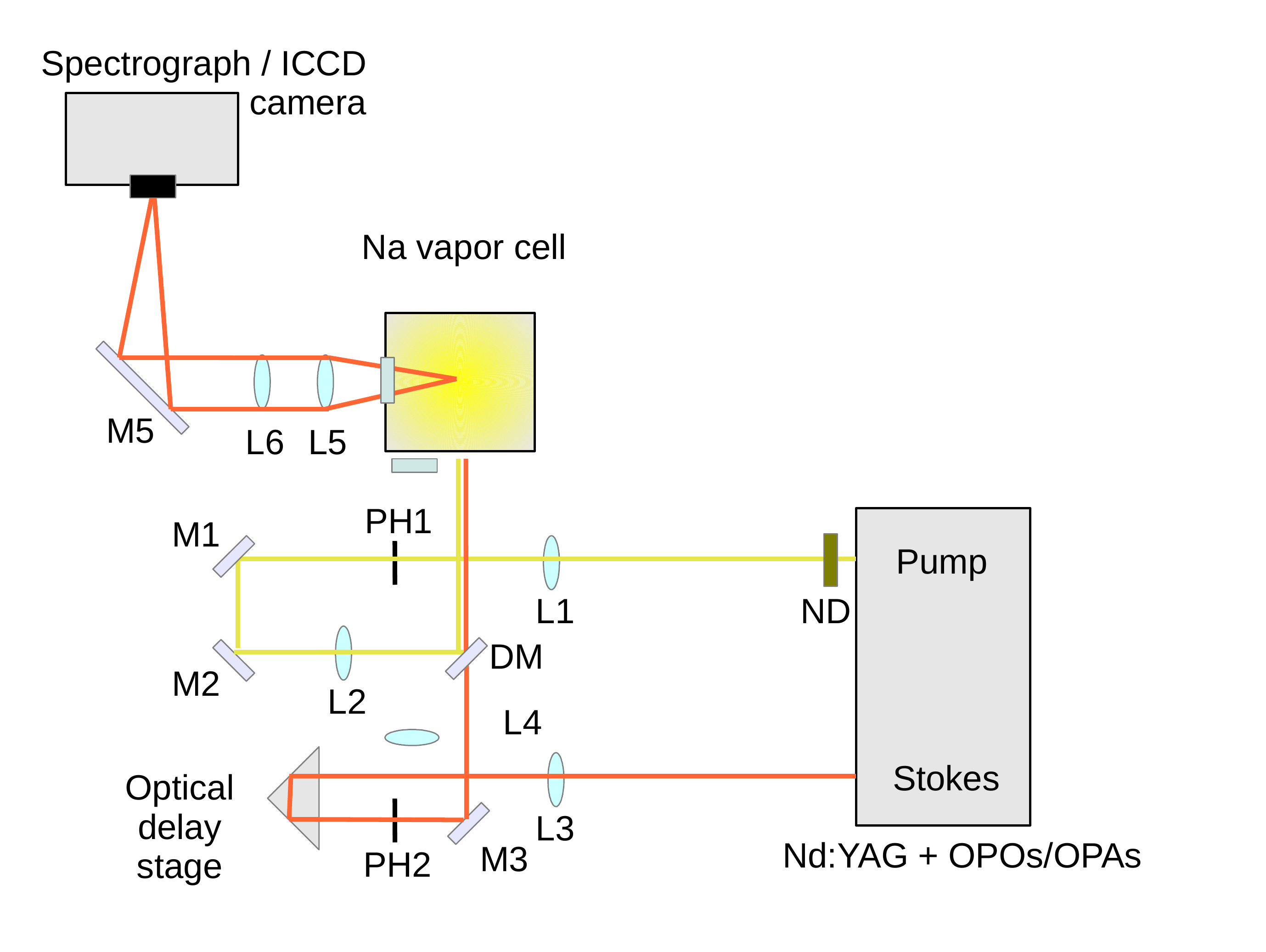}
\caption{\label{fig:experiment} (Color online) A diagram of our experimental arrangement. ND: neutral density filter (1.0-1.5 OD); L1, L2, L3, L4, L5, and L6: lenses of focal lengths 1000 mm, 750 mm, 1000 mm, 450 mm, 203 mm, and 203 mm respectively; M1, M2, M3, M4, M5: mirrors; PH1 and PH2: ceramic pinholes of diameter 400 $\mu$m; DM: dichroic mirror.}
\end{figure}

Fig.~\ref{fig:experiment}  gives a diagram of the experimental arrangement. Pump and Stokes laser pulses were generated by two synchronously-pumped OPO/OPA units (EKSPLA PG 411) at a repetition rate of 10 Hz. A streak camera (Hamamatsu C7700) was used to determine the pulse widths (See Table~\ref{tab:pulse_parameters}.) and relative timing of the pump and Stokes laser pulses. The wavelengths (in air) of the resonant pump and Stokes pulses were 589.59 nm and 615.42 nm respectively for the D1 channel and 588.99 nm and 616.07 nm respectively for the D2 channel as measured by a HighFinesse WS5 wavelength meter. The linewidths of the pump and Stokes lasers were measured by a Fabry-P\'{e}rot interferometer \cite{Johnson2011} and time-bandwidth products were calculated. (See Table~\ref{tab:pulse_parameters}.) The beams were focused onto ceramic pinholes for spatial filtering; the collimated spatial pulse profile was then determined by projecting the pulse onto a flat surface and measuring the intensity with a CCD camera. The intensity profiles of the beams displayed several rings of an Airy pattern and the central peaks were fit with Gaussians. The beam radii of the pump and Stokes pulses are given in Table~\ref{tab:pulse_parameters}. 

The timing of the Stokes pulse relative to the pump pulse was adjusted using an optical delay line driven by a Newport LTA-HS (0.035 $\mu$m resolution) and measured with a streak camera. The error in the temporal overlap is estimated to be within $\pm$ 2.0 ps. Since the light that pumps the OPO/OPA units originates from the same pulse from a mode-locked Nd:YVO$_4$ oscillator, the timing jitter between the pulses is negligible. Additional details about the laser system have been described previously \cite{Johnson2010a}.

\begin{table}[b]
\caption{\label{tab:pulse_parameters}%
Measured pulse parameters for the pump and Stokes beams. 
}
\begin{ruledtabular}
\begin{tabular}{lcc}
\textrm{Pulse parameters}&
\textrm{Pump}&
\textrm{Stokes}\\
\colrule
Radius, $W$\footnote{Radius where the irradiance drops to $1/e^2$ of its value at beam center.} (mm) &0.79(2) & 0.70(5) \\
Pulse width\footnote{Full width at half maximum (FWHM). The pulse broadening from the streak camera was corrected.} (ps) & 7.3(4) & 8.6(23) \\
$\Delta \tau$\footnote{Time when the irradiance drops to $1/e^2$ of the maximum.} (ps) & 6.2(3) & 7.3(19) \\
Linewidth\footnote{FWHM} (GHz) & 70(21) & 67(12)\\
TBP\footnote{Time-bandwidth product. Values are given as a multiple of 0.4413, the TBP for Fourier-transform-limited Gaussian pulses.}& 1.1(3) & 1.3(2) \\
\end{tabular}
\end{ruledtabular}
\end{table}

Population transfer was detected by monitoring the fluorescence doublet from the 5s state to the 3p states (616.07 and 615.42 nm). The fluorescence signal was collected perpendicular to the laser propagation direction and imaged onto the entrance slit of a spectrograph (Princeton Instruments/Acton SpectraPro 2300i) with f/4 optics. The entrance slit of the spectrograph was set at 28 $\mu$m. Fluorescence was detected with an ICCD camera (Andor iStar). To suppress Rayleigh scattering from the laser, the gate of the ICCD camera was adjusted to begin collecting light a few ns after the Rayleigh scattered light had disappeared and integrated the fluorescence for 1.0 $\mu$s. In the results that follow, 100 laser shots were averaged for each data point. Measurements of the pulse-to-pulse energy stability for the pump and Stokes pulses yielded a standard deviation of 15($\pm$5)\%. 

Sodium vapor was produced from pure metallic sodium inside a heated vapor cell.  Both glass and stainless steel vapor cells were used. In the details that follow, the elevated cell temperatures were solely for the purpose increasing the vapor pressure of sodium. The glass cell was purchased from Precision Glassblowing (Colorado). The glass cell is T-shaped ($7.62\, {\rm cm} \times 2.54 \, {\rm cm} \times 5.08 \, {\rm cm}$) with three fused-silica windows, two in-line for the laser entrance and exit and one perpendicular to collect the fluorescence. The vapor pressure of the sodium was controlled by heating the cell inside an oven. The glass cell was heated to $120 \,^{\circ}\mathrm{C}$, yielding a sodium vapor pressure of $10^{-6}$ torr. This was the highest temperature to which we could elevate the cell without the sodium turning the glass cell brown. The stainless steel vapor cell was constructed from a 304L stainless steel cube (CF 275-CU, A\&N Corporation) with ConFlat seals. The cell had  two fused-silica windows on opposite sides for laser transmission and a sapphire window for fluorescence collection. The stainless steel cell was housed in an oven with two windows to allow the laser light to pass through and a window perpendicular to the laser propagation direction for collection of fluorescence. The cell had an extended side arm and gate valve that protruded outside the oven and which were wrapped with heating tape. The gate valve led to a turbomolecular pump allowing for high vacuum evacuation. The temperature of the side arm was held in the range of $175 - 205 \,^{\circ}\mathrm{C}$ which was approximately $30\,^{\circ}\mathrm{C}$ cooler than the cube ($205 - 235 \,^{\circ}\mathrm{C}$). The cooler side arm prevented sodium from accumulating on the windows of the cube. The temperature of the cube was limited by the windows: higher temperatures caused the windows to lose their vacuum seal. 

\section{Calibration}

\subsection{\label{sec:beam_parameters} Beam parameters}

Stimulated emission pumping (SEP) can be used as a STIRAP calibration method. SEP is implemented with a pulse ordering opposite that of STIRAP; the Stokes pulse follows the pump pulse. The pump pulse transfers sodium from the 3s ground state to one of the 3p intermediate states; the Stokes pulse transfers the sodium from the 3p state to the 5s final state. Unlike STIRAP, SEP can be achieved with coherent or incoherent light. When it is coherent (as is the case with the lasers used in this experiment), Rabi oscillations occur in which the pump pulse completely transfers the sodium population back and forth between the initial and intermediate states. Since the energy of the pulse varies with position within the beam, the population of the intermediate state of sodium also varies. For large integrated Rabi frequencies, many oscillations occur at the center of the pulse with correspondingly fewer oscillations radially outward. The population of the intermediate state thus oscillates multiple times between 0 and 100\% in moving radially outward from the center of the pulse with the average tending toward 50\%. The Stokes pulse does the same with the population in the intermediate state transferring an average of 50\% of the 50\% in the intermediate state to the final state, yielding 25\% in the 5s state. The STIRAP efficiency can therefore be checked by comparison with the SEP efficiency. 

With 100\% STIRAP efficiency, the STIRAP fluorescence signal may be expected to be $4 \times$ larger than the SEP fluorescence. However, quantum interference reduces the maximum expected efficiency below 100\% (Sec.~\ref{sec:STIRAP_in_sodium}). In addition, since the spectrograph that collects the fluorescence signal collects light from the intense center of the beam as well as from the edges of the beam (where SEP efficiency is small and STIRAP efficiency is even smaller), the measured STIRAP efficiency is reduced further. (As can be seen with the analysis in Appendix~\ref{sec:modelfluor}, the reduction in efficiency due to the spectrograph happens even with an infinitesimal entrance slit width.) 

The reduction in efficiency described above was determined using the measured values of the pulse energy $U(E_0)$ (with $\pm$10-15\% fluctuation), pulse width $\Delta \tau$, and beam radius $W$ to calculate the Rabi frequencies as a function of position within the beam. These Rabi frequencies were used to calculate system Hamiltonians which were solved to find the transfer efficiencies of STIRAP and SEP processes as a function of position within the beam. The STIRAP/SEP ratio was then calculated and compared to measured data. Details for the calculation are provided in Appendix~\ref{sec:modelfluor}.

\begin{figure}[t]
\includegraphics[width=0.95\linewidth]{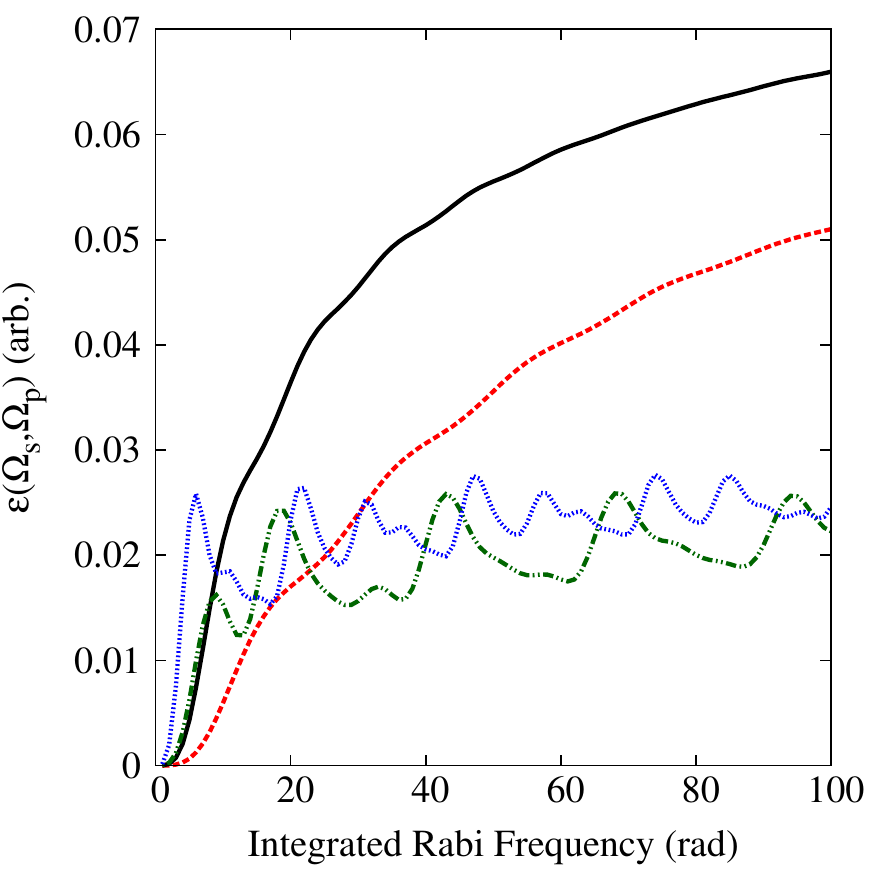}\\
\includegraphics[width=0.90\linewidth]{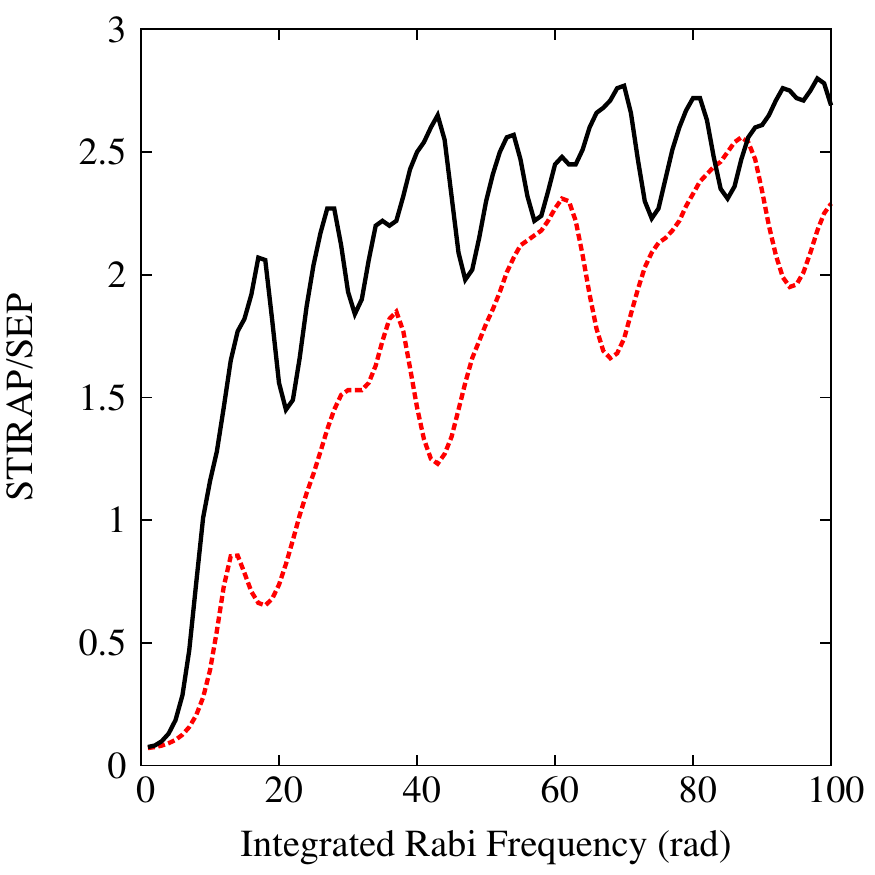}
\caption{\label{fig:STIRAPSEP} (Color online) Top: Calculations of the quantity of light collected by the spectrograph ($\epsilon$ described in Appendix~\ref{sec:modelfluor}), as a function of integrated Rabi frequency of the Stokes pulse, for D1 STIRAP (dashed red line), D2 STIRAP (solid black line), D1 SEP (dotted blue line) and D2 SEP (dash-dotted green line) transitions. Bottom: The calculated STIRAP to SEP ratios, as a function of integrated Rabi frequency of the Stokes pulse, for the D1 (dashed red line) and D2 (solid black line) transitions.}
\end{figure}

In Fig.~\ref{fig:STIRAPSEP}, the upper panel shows a calculation of the quantity of light that enters the spectrograph as a function of integrated Rabi frequency for the Stokes pulse (calculated with Eq.~\ref{eq:int_rabi_freq}). The calculation was performed at a Stokes to pump integrated Rabi frequency ratio that matches the experimental conditions in Section~\ref{sec:observations} (See Figs.~\ref{fig:STIRAP_delay_scan_SP} and \ref{fig:STIRAP_delay_scan_SS}), i.e. 13.7 rad (pump) and 33.1 rad (Stokes) for the transition through the $3\,^2{\rm P}_{1/2}$ state and 19.4 rad (pump) and 34.3 rad (Stokes) for the transition through the $3\,^2{\rm P}_{3/2}$ state. The figure displays STIRAP transitions through the $3\,^2{\rm P}_{1/2}$ and $3\,^2{\rm P}_{3/2}$ (the dashed red and solid black lines, respectively) and the corresponding quantities for the SEP transitions through the $3\,^2{\rm P}_{1/2}$ and $3\,^2{\rm P}_{3/2}$ (the dotted blue and dash-dotted green lines, respectively). 

The bottom panel in Fig.~\ref{fig:STIRAPSEP} shows the STIRAP to SEP ratio of the data shown in the top panel, again relative to the integrated Rabi frequency of the Stokes pulse. Transitions tuned through the $3\,^2{\rm P}_{1/2}$ and $3\,^2{\rm P}_{3/2}$  states are the dashed red and solid black lines, respectively. The data shown in the bottom two panels of Fig.~\ref{fig:STIRAP_delay_scan_SS} indicate a STIRAP to SEP ratio of 1.6 for the transition through $3\,^2{\rm P}_{1/2}$ (lower-left panel) and 2.3 for the transition through $3\,^2{\rm P}_{3/2}$ (lower-right panel). This is in good agreement with predicted values from the bottom panel of Fig.~\ref{fig:STIRAPSEP} of 1.6 through the $3\,^2{\rm P}_{1/2}$ state and 2.13 through the $3\,^2{\rm P}_{3/2}$ state at a Stokes integrated Rabi frequency of 34 rad. 

While these efficiencies appear low, the analysis here together with Appendix~\ref{sec:modelfluor} demonstrate that they reflect the maximum expected STIRAP efficiency of 80\% at the center of the beam (reduced from 100\% due to quantum interference between excitation pathways through the $3\,^2{\rm P}_{1/2}$ and $3\,^2{\rm P}_{3/2}$ states). 

\begin{figure*}
\includegraphics[width=3.5in]{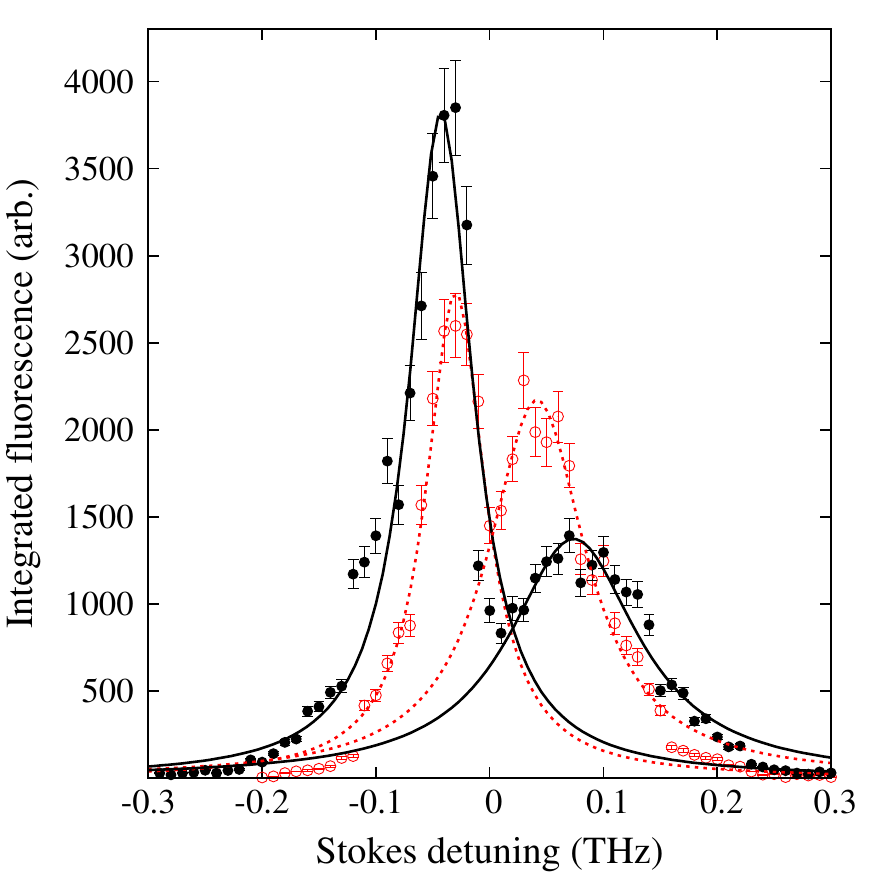}
\includegraphics[width=3.5in]{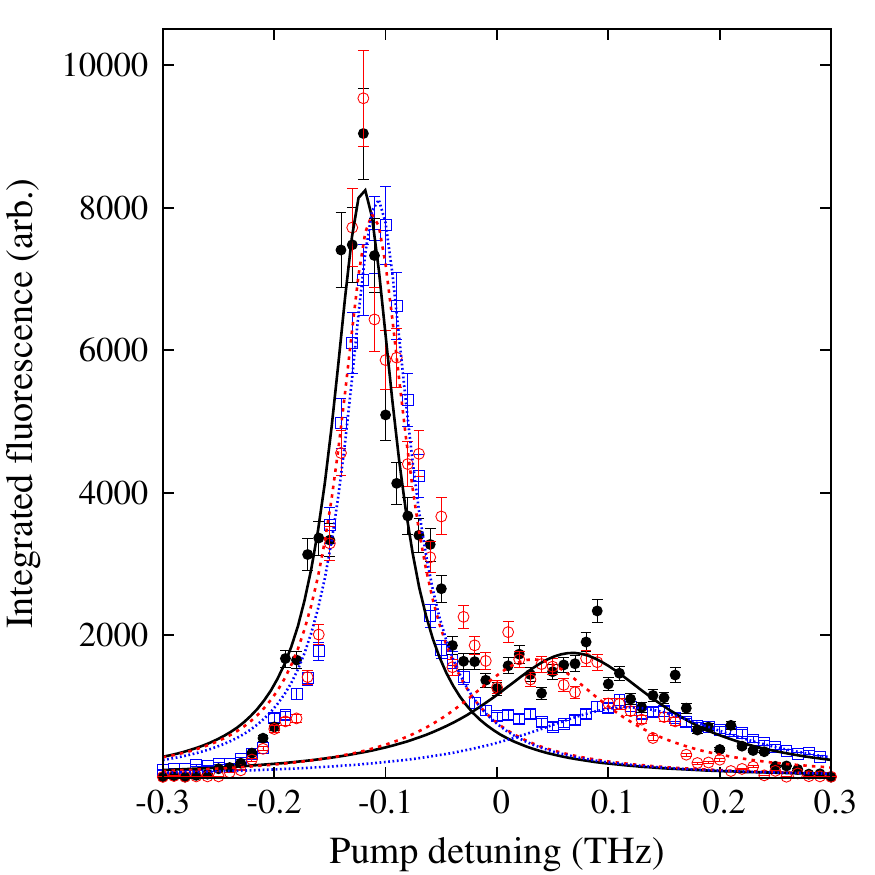}\\
\includegraphics[width=2.3in]{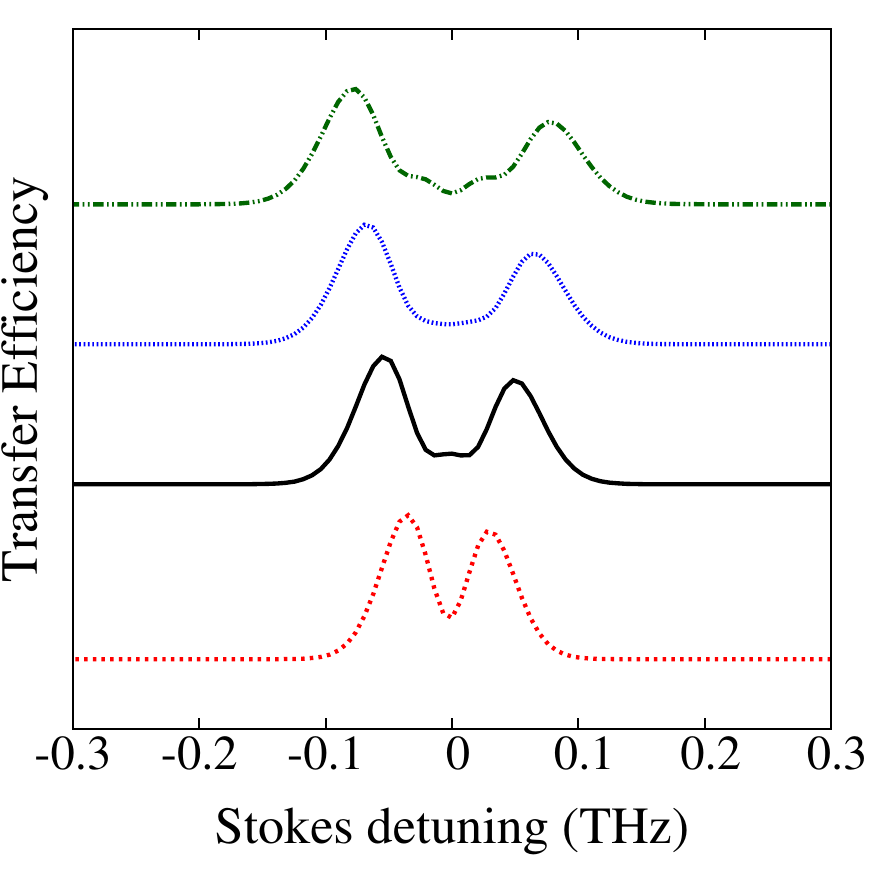}
\includegraphics[width=2.3in]{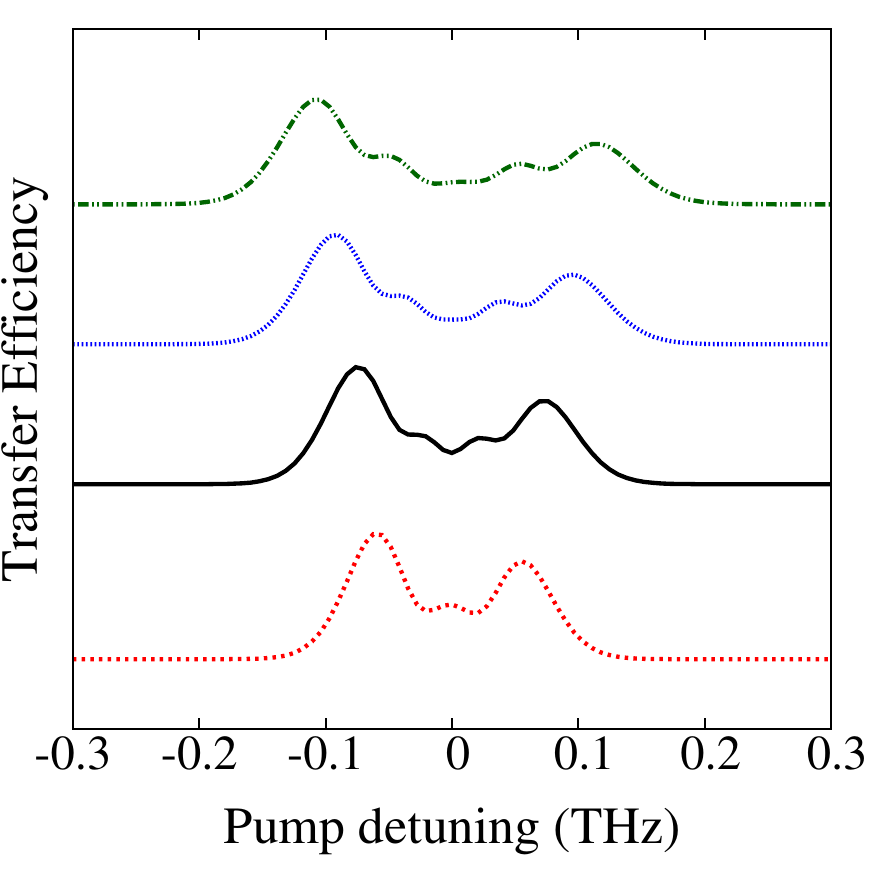}
\includegraphics[width=2.3in]{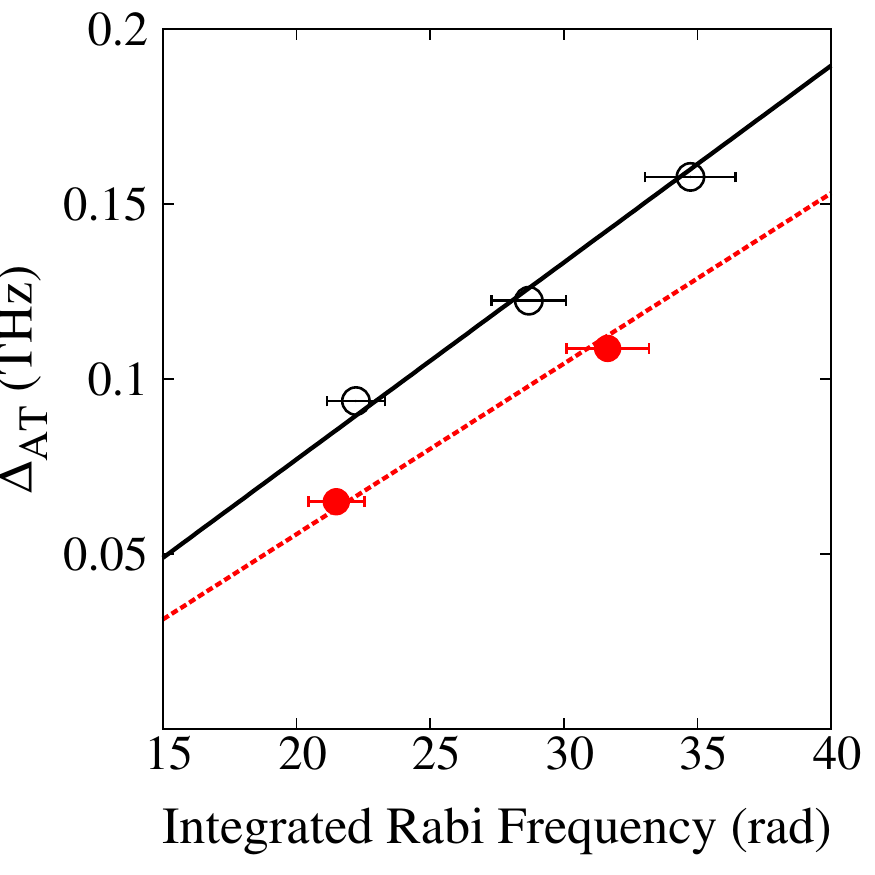}
\caption{\label{fig:AutlerTownes} (Color online) Top left: Data and Lorentzian fits for Autler-Townes (AT) splitting with strong pump ($3\,^2{\rm P}_{3/2}$ $\leftrightarrow$ 3s transition) and weak Stokes for two pump energies: 0.46 and 1.0 $\mu$J (Fluence: 0.047 and 0.10 mJ/cm$^2$; Integrated Rabi frequency: 21.4 and 31.5 rad) for the dashed-red and solid-black plots, respectively. Top right: Data and Lorentzian fits for AT splitting with strong Stokes (5s $\leftrightarrow 3\,^2{\rm P}_{3/2}$) and weak pump for three Stokes energies: 18, 30, and 44 $\mu$J (Fluence: 2.3, 3.9, and 5.7 mJ/cm$^2$; Integrated Rabi frequency: 22.2, 28.7, and 34.7 rad) for the dashed-red, solid-black and dotted-blue plots, respectively. Bottom left: Transfer efficiencies calculated to show Autler-Townes splitting ($\Delta_{AT-SP}$) for strong pump energies of 0.5, 1.0, 1.5 and 2.0 $\mu$J (Fluence: 0.051. 0.10, 0.15, and 0.20 mJ/cm$^2$; Integrated Rabi frequency: 22.3, 31.5, 38.6, and 44.5 rad) from bottom to top. The curves are offset vertically for better visibility. Bottom center: Transfer efficiencies calculated to show Autler-Townes splitting ($\Delta_{AT-SS}$) for strong Stokes energies of 20, 30, 45 and 60 $\mu$J (Fluence: 2.6, 3.9, 5.8, and 7.8 mJ/cm$^2$; Integrate Rabi frequency: 23.4, 28.7, 35.1, and 40.5 rad) from bottom to top. Bottom right:Autler-Townes splitting for theory and data. The dashed red line and closed circles represent theory and data for the strong pump/weak Stokes case, and the solid black line and open circles represent theory and data for the strong Stokes/weak pump case.}
\end{figure*}

\subsection{Autler-Townes}

The Autler-Townes effect can be used to calibrate the parameters of the lasers used in this experiment. The Autler-Townes (AT) splitting for the sodium D2 resonance was measured (5s $\leftarrow 3\,^2{\rm P}_{3/2}\leftarrow$ 3s), and then compared to the expected value of the splittings calculated by the system's Hamiltonian. The splitting was measured for both Stokes (5s $\leftrightarrow 3\,^2{\rm P}_{3/2}$) and pump (3\,$^2{\rm P}_{3/2} \leftrightarrow$ 3s) splittings, probed by the relatively weak pump and Stokes pulses, respectively.

The upper left panel of Fig.~\ref{fig:AutlerTownes} shows an example of two measured AT-splittings of the $3\,^2{\rm P}_{3/2}$ level for strong pump and relatively weak Stokes pulses. The lines (solid and dashed) in both upper panels are fits with two Lorentzian peaks calculated by a Levenberg-Marquardt algorithm. The red data with corresponding dashed curve represent a pump beam energy 0.46 $\mu$J, corresponding to an integrated Rabi frequency of 21.4 rad, which splits the $3\,^2{\rm P}_{3/2}$ level by 0.065 THz (determined by the positions of the peaks in the fit), and the black data with corresponding solid curve represent an energy of 1.0 $\mu$J (31.5 rad), which splits the $3\,^2{\rm P}_{3/2}$ level by 0.109 THz. The upper right panel is a similar plot of the AT-splittings of the $3\,^2{\rm P}_{3/2}$ level for strong Stokes and relatively weak pump pulses. The red data with corresponding dashed curve represent a Stokes energy of 18 $\mu$J (22.2 rad), which splits the $3\,^2{\rm P}_{3/2}$ level by 0.109 THz. The black data with corresponding solid curve represent a Stokes energy of 30 $\mu$J (28.7 rad), which splits the $3\,^2{\rm P}_{3/2}$ level by 0.142 THz. The blue data with corresponding dotted curve represent a Stokes energy of 44 $\mu$J (34.7 rad), which splits the $3\,^2{\rm P}_{3/2}$ level by 0.183 THz. The significantly smaller error bars for the 44 $\mu$J pulses (blue squares) are attributed to having approximately an order of magnitude more counts for the integrated fluorescence when compared to the 18 and 30 $\mu$J data. 

In the bottom left panel of Fig.~\ref{fig:AutlerTownes}, the strong-pump splitting, $\Delta_{AT-SP}$, is calculated from the Hamiltonian for four pump pulse energies (0.5, 1.0, 1.5, and 2.0 $\mu$J, from bottom to top). In the bottom middle panel of Fig.~\ref{fig:AutlerTownes}, the strong-Stokes splitting $\Delta_{AT-SS}$ is calculated from the Hamiltonian for four Stokes energies (20, 30, 45, and 60 $\mu$J, from bottom to top). In these calculations, we take into account the spatial Gaussian nature of the pump and Stokes pulses, and average the AT-splitting over the entire envelope of the pulse as described in Appendix~\ref{sec:modelfluor}. 

The bottom right panel of Fig.~\ref{fig:AutlerTownes} displays theoretical calculations (dashed red line) and data (solid red circles) for strong pump and weak Stokes AT splittings and theoretical calculations (solid black line) and data (hollow black circles) for weak pump and strong Stokes AT splittings. The data in this panel is that shown in the top left and right panels of this figure. The theoretical calculations and data are in good agreement. The uncertainty in beam energies (horizontal error bars) is produced by the energy jitter given in Section~\ref{sec:beam_parameters}.
 
\section{STIRAP experiments and simulations}

\subsection{STIRAP vs. timing delay\label{sec:observations}}

\begin{figure*}
\includegraphics[width=3.4in]{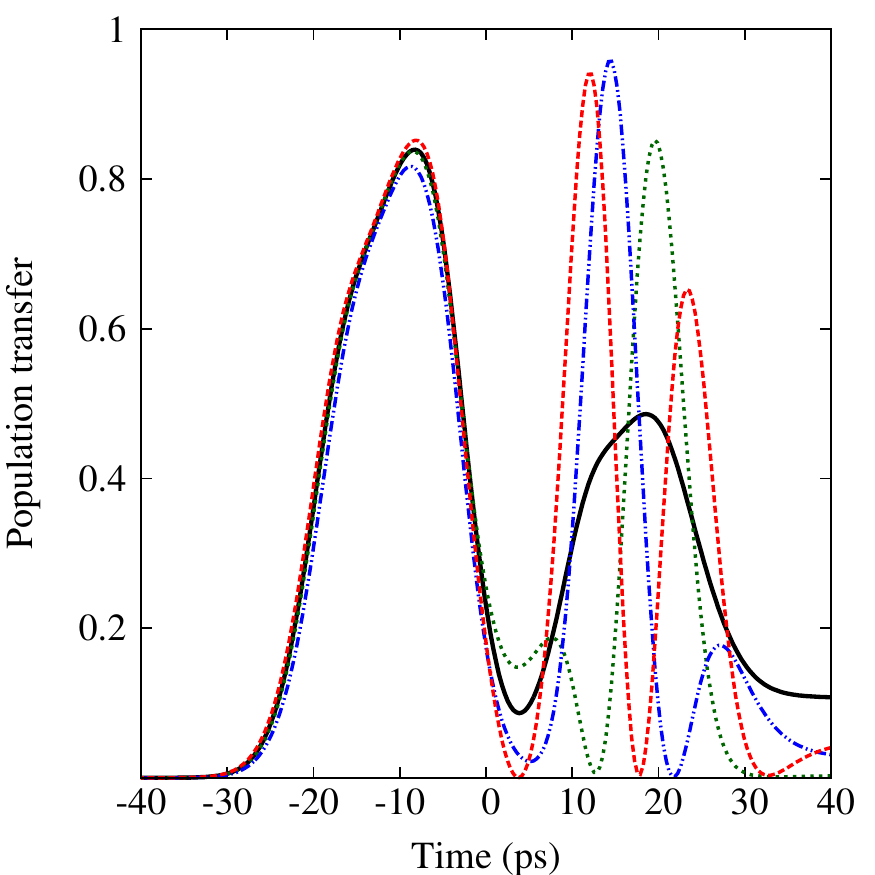}
\includegraphics[width=3.4in]{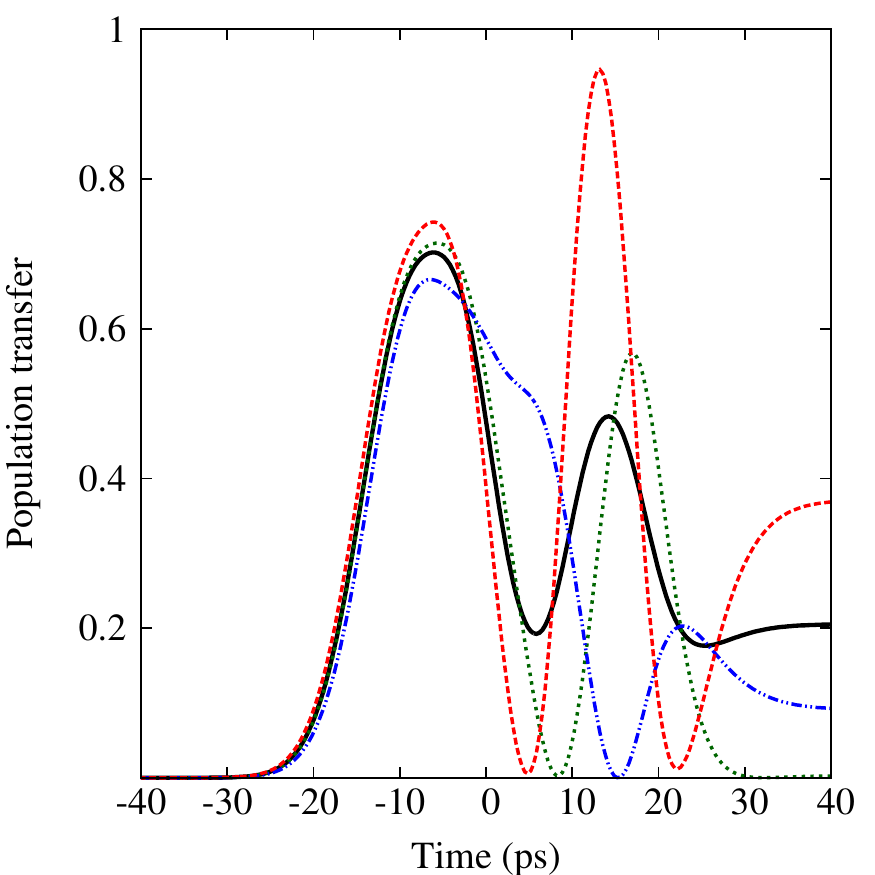}\\
\includegraphics[width=3.5in]{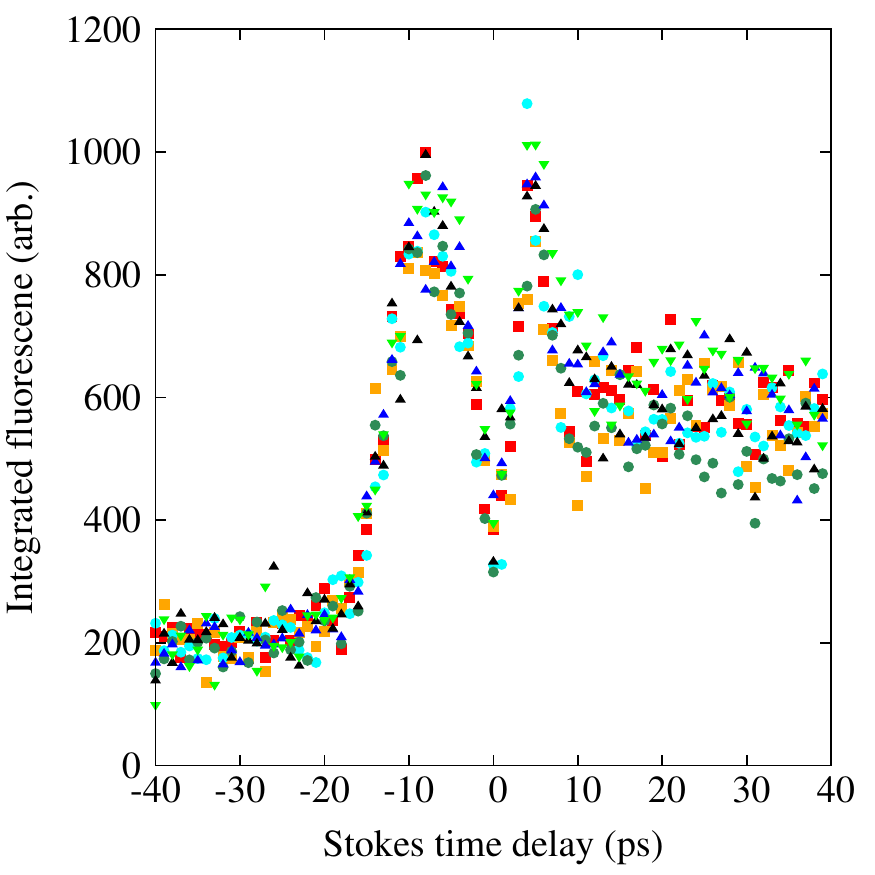}
\includegraphics[width=3.5in]{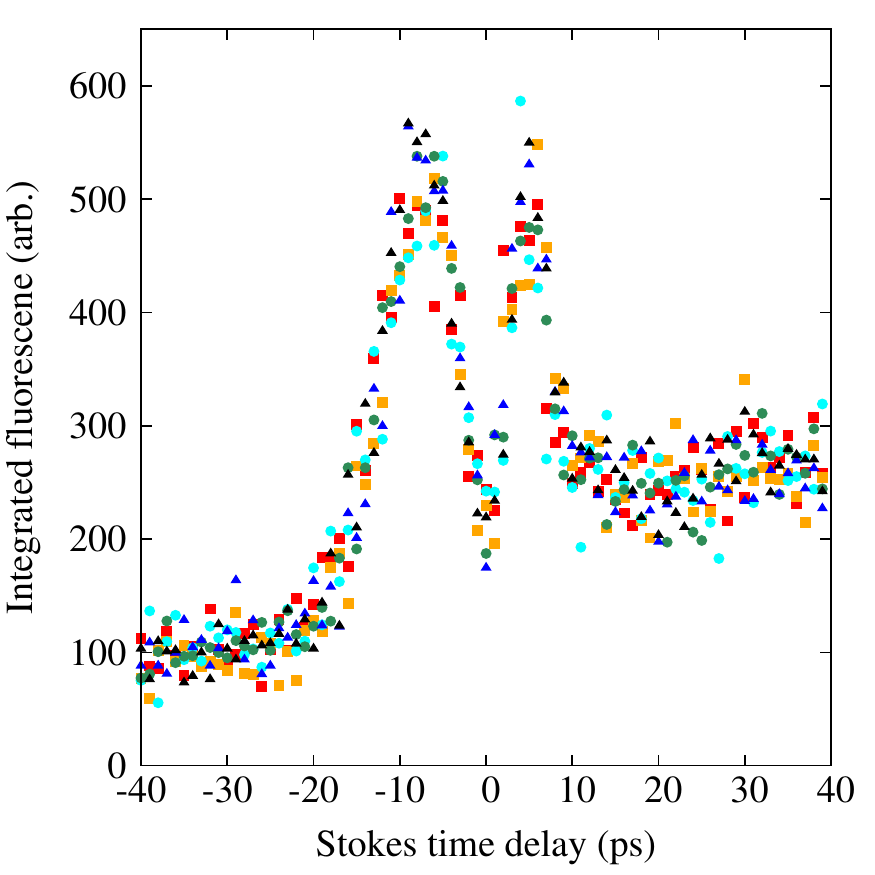}
\caption{\label{fig:STIRAP_delay_scan_SP} (Color online) {\bf Strong Pump} Fluorescence from the 5s state to the 3p states as a function of timing delay between the pump and Stokes pulses for the D1 (left) and D2 (right) channels with the pump pulse stronger than the Stokes pulse. {\bf Top (computation):} Theoretical transfer efficiencies calculated for both the D1 and D2 channels, where the solid black line represents the predicted population transfer for beam energies averaged around a Gaussian model for the beam jitter with a width of $\pm$15\%, the green dotted line represents a calculation with beam energies from the panels directly below, and the red dashed and blue dot-dashed lines represent calculations with energies 15\% above and below those of the green dotted line. {\bf Bottom (data):} The pump and Stokes pulse energies for the D1 channel are ${\rm E_{pump}} = 15(1) \,\mu {\rm J}$ (Fluence: 1.5 mJ/cm$^2$, Integrated Rabi frequency: 86.3 rad) and ${\rm E_{Stokes}} = 17(2)  \,\mu {\rm J}$ (Fluence: 2.23 mJ/cm$^2$, Integrated Rabi frequency: 21.6 rad).  For the D2 channel ${\rm E_{pump}}  = 6.6(3)  \,\mu {\rm J}$ (Fluence: 0.67 mJ/cm$^2$, Integrated Rabi frequency: 81.0 rad) and ${\rm E_{Stokes}} = 4.4(3)  \,\mu {\rm J}$ (Fluence: 0.57 mJ/cm$^2$, Integrated Rabi frequency: 11.0 rad). In the bottom two panels, different symbols are used to display multiple data sets. Negative Stokes time delay indicates STIRAP pulse order.}
\end{figure*}

\begin{figure*}
\includegraphics[width=3.4in]{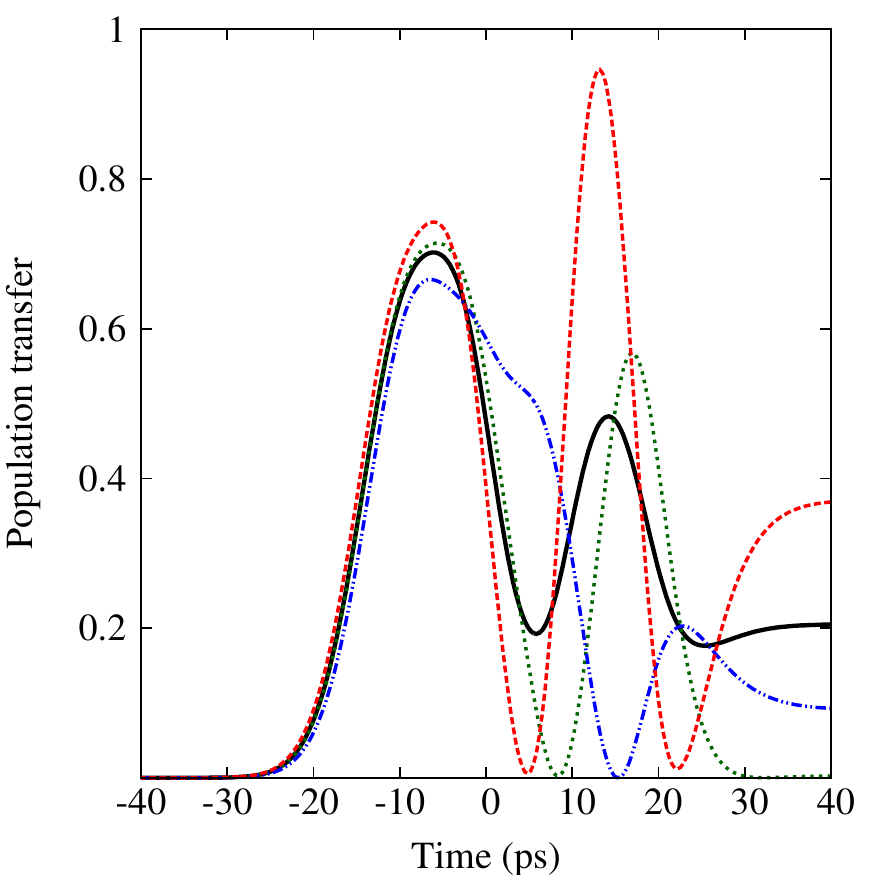}
\includegraphics[width=3.4in]{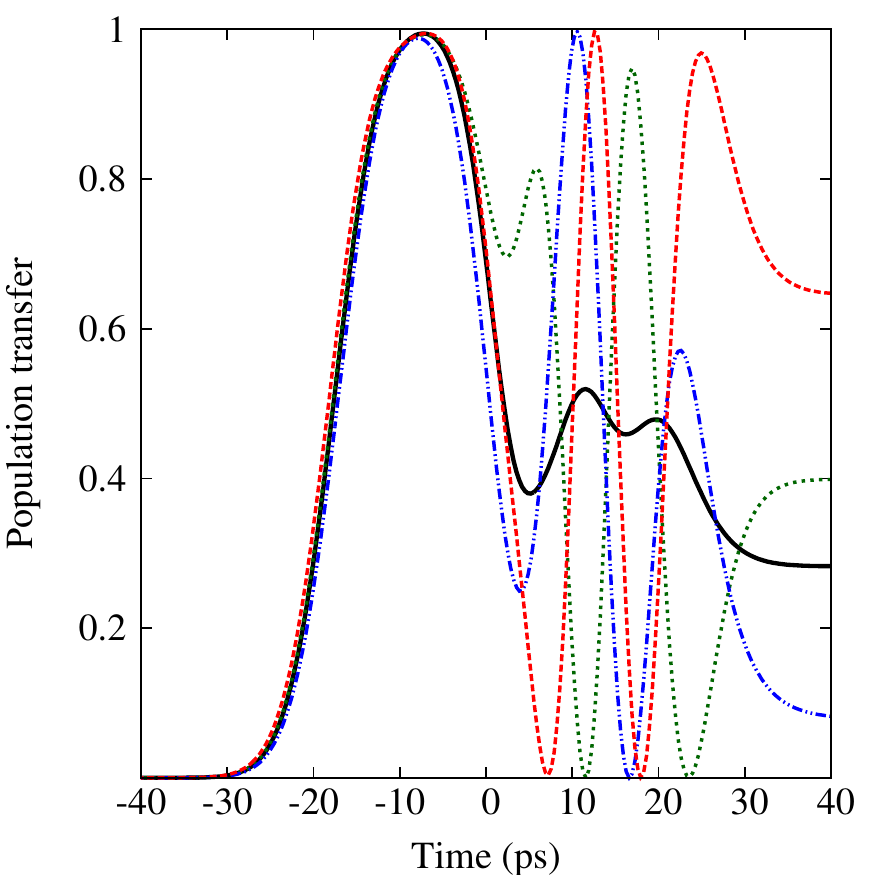}\\
\includegraphics[width=3.5in]{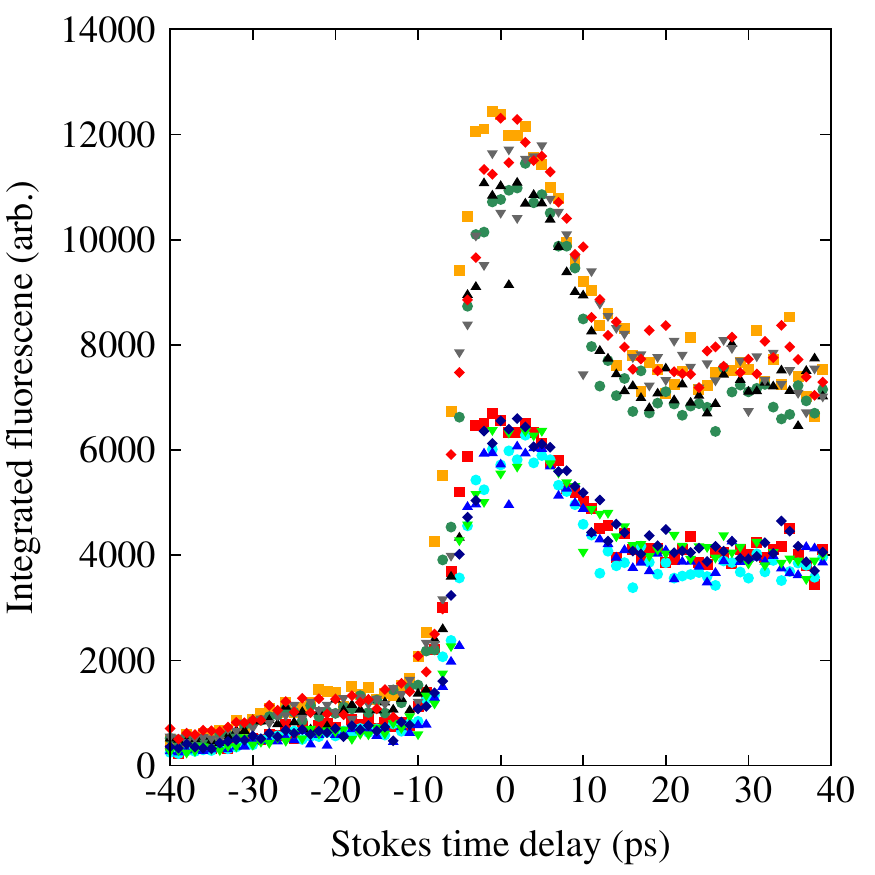}
\includegraphics[width=3.5in]{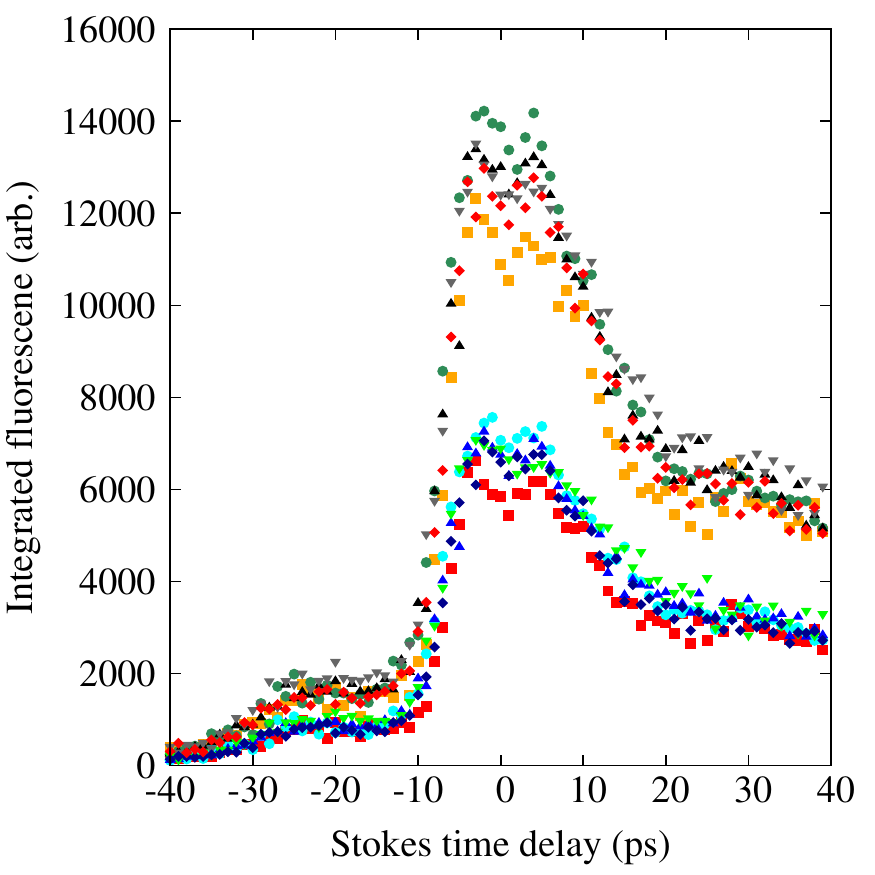}
\caption{\label{fig:STIRAP_delay_scan_SS} (Color online) {\bf Strong Stokes} Fluorescence from the 5s state to the 3p states as a function of timing delay between the pump and Stokes pulses for the D1 (left) and D2 (right) channels with the Stokes pulse stronger than the pump pulse. {\bf Top (computation):} See caption to Fig.~\ref{fig:STIRAP_delay_scan_SP}. {\bf Bottom (data):} The energies for the pump and Stokes pulses for the D1 channel are ${\rm E_{pump}} = 0.38(6) \,\mu {\rm J}$ (Fluence: 0.039 mJ/cm$^2$, Integrated Rabi frequency: 13.7 rad) and ${\rm E_{Stokes}} = 40(6)\,\mu {\rm J}$ (Fluence: 5.2 mJ/cm$^2$, Integrated Rabi frequency: 33.1 rad). For the D2 channel ${\rm E_{pump}} = 0.38(6)  \,\mu {\rm J}$ (Fluence: 0.039 mJ/cm$^2$, Integrated Rabi frequency:19.4 rad) and ${\rm E_{Stokes}} = 43(5) \,\mu {\rm J}$ (Fluence: 5.6 mJ/cm$^2$, Integrated Rabi frequency: 34.3 rad). Fluorescence from the 5s to the two 3p states were measured independently. Fluorescence from 5s $\rightarrow$ $3\,^2{\rm P}_{3/2}$  is 2$\times$ larger than from 5s $\rightarrow$ $3\,^2{\rm P}_{1/2}$. }
\end{figure*}

Typical plots of the transfer efficiency from the 3s to the 5s state as a function of the timing delay between the pump and Stokes pulses are shown in Figs.~\ref{fig:STIRAP_delay_scan_SP} and \ref{fig:STIRAP_delay_scan_SS}. Figure~\ref{fig:STIRAP_delay_scan_SP} displays theory and data when the pump pulse is stronger than Stokes pulse; Fig.~\ref{fig:STIRAP_delay_scan_SS} displays theory and data when the Stokes pulse is stronger than the pump pulse. In each of these figures, the top two panels are solutions to Eq.~\ref{eq:fine} with pump and Stokes energies and temporal pulse profiles that correspond to the data in the panels immediately below. In both Figs.~\ref{fig:STIRAP_delay_scan_SP} and \ref{fig:STIRAP_delay_scan_SS} the left two panels display theory and data for excitation through the D1 channel and the right two panels for excitation through the D2 channel. 

In Figs.~\ref{fig:STIRAP_delay_scan_SP} and \ref{fig:STIRAP_delay_scan_SS} there is generally good agreement between theory and data. All computations and data display prominent STIRAP peaks for negative Stokes time delays. It is interesting to note that in Fig.~\ref{fig:STIRAP_delay_scan_SP} theory predicts and data confirms the existence of a persistent dip in the transfer efficiency when the Stokes and pump pulses are coincident (zero Stokes time delay). This dip is reminiscent of electromagnetically induced transparency (EIT) albeit with the pump pulse serving as the coupling field between levels $|1\rangle$ and $|2\rangle$, opposite the more common EIT case with strong coupling between $|2\rangle$ and $|3\rangle$. In contrast, for Fig.~\ref{fig:STIRAP_delay_scan_SS} the theoretical calculations at three energies near the energy of the data display large oscillations in the signal that tend to average out for zero and positive Stokes time delays (SEP). For both Figs.~\ref{fig:STIRAP_delay_scan_SP} and \ref{fig:STIRAP_delay_scan_SS} the fact that some peaks persist in the theoretically calculated average, suggests that it is likely that for this data, laser energy fluctuations were larger than listed in Sec.~\ref{sec:exp_setup} which caused SEP peaks to disappear in the data. 

The location of the STIRAP peaks in Figs.~\ref{fig:STIRAP_delay_scan_SP} and \ref{fig:STIRAP_delay_scan_SS} are located at negative Stokes time delays, however the STIRAP peak location for the data in the lower panels are generally not as negative as predicted by theory in the upper panels. The cause of the discrepancy is not clear although it may be due to the lasers operating near, but not at the Fourier transform limit.

\subsection{STIRAP ridge} 

\begin{figure}[t]
\includegraphics[width=3.5in]{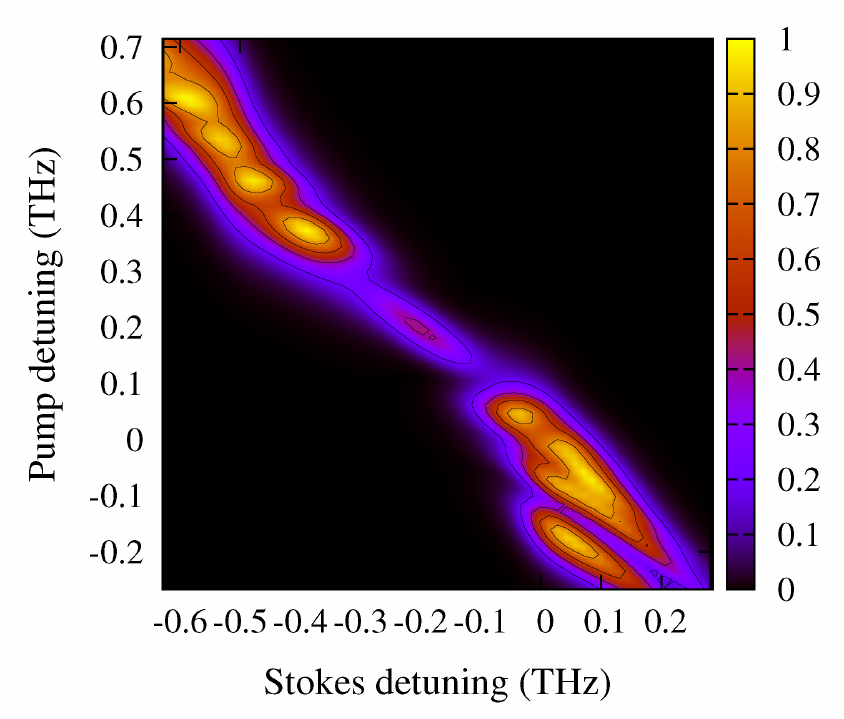}\\
\vspace{-3.8em}
\includegraphics[width=3.5in]{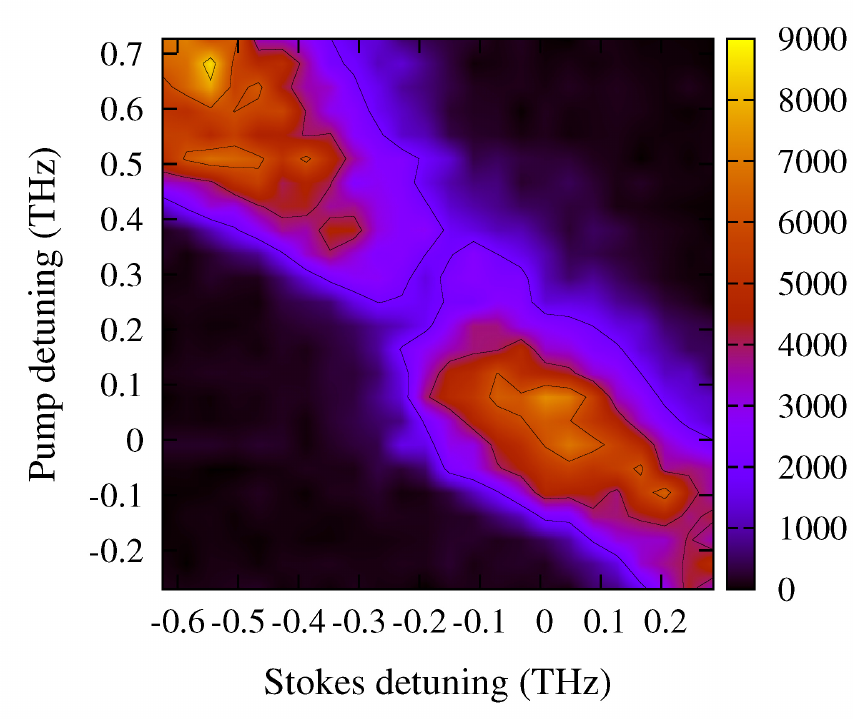}
\caption{\label{fig:STIRAP_ridge_expt1} (Color online) {\bf Top:} Fluorescence (arbitrary units) from the 5s to the 3p states as a function of the frequencies of the pump and Stokes beams. Zero detuning corresponds to excitation through the D1 channel. The temporal overlap between pump and Stokes pulses was fixed in the STIRAP orientation with the Stokes preceding the pump by 5.0 ps. The measured pulse duration for the pump and Stokes pulses was 8(1) ps. {\bf Bottom:} Computational prediction of STIRAP efficiency for the same setup as the top figure, for pump pulse of 6 $\mu J$ (Fluence: 0.61 mJ/cm$^2$; Integrated Rabi frequency: 56.8 and 77.3 rad for D1 and D2 transitions, respectively) and a Stokes pulse of 220 $\mu J$ (Fluence: 28.6 mJ/cm$^2$; Integrated Rabi frequency: 83.1 and 77.7 rad for D1 and D2 transitions, respectively).}
\end{figure}

In ordinary three-state STIRAP a two-dimensional plot versus the frequency detuning of the pump and Stokes beams reveals a ridge along which the transfer efficiency remains high. In ladder STIRAP the efficiency remains high where there is a two-photon resonance; the sum of the energies of the pump and Stokes photons is equal to the difference in energy between the initial and final states.

The result of a 2-dimensional frequency scan (of pump and Stokes beams) in the STIRAP order for sodium vapor is shown in Fig.~\ref{fig:STIRAP_ridge_expt1}. Unlike the ridge seen in three-state STIRAP, in this plot there are two distinct STIRAP islands. The transfer efficiency is greatest when the pump and Stokes lasers are tuned to resonance through the $3\,^2{\rm P}_{3/2}$ channel (upper-left island). The smaller island near the center of the plot occurs when the lasers are tuned to resonance through the $3\,^2{\rm P}_{1/2}$ state. The dip between the two islands displays evidence for destructive interference between the pathways through the two 3p states. 

\section{Conclusion}

We have demonstrated STIRAP in a gas of sodium atoms above room temperature using picosecond lasers. We show that both 3p states contribute to the transfer efficiency revealing evidence for quantum interference between alternate pathways. The Autler-Townes data presented display shifts that are in agreement with theoretical predictions. The efficiency of the STIRAP transfer process is near 100\% at the center of the beams. A two-dimensional scan of the pump and Stokes wavelengths display a STIRAP ridge with a valley between the resonances through the $3\,^2{\rm P}_{1/2}$ and $3\,^2{\rm P}_{3/2}$ islands of high efficiency.

Future work could explore the rich interference region between the two islands displayed in Fig.~\ref{fig:STIRAP_ridge_expt1}. Preliminary computations reveal the existence of a series of ridges of high and low transfer efficiency with increasing integrated Rabi frequencies that have yet to be explored experimentally.

\begin{acknowledgments}
We are grateful to the United States Army RDECOM Night Vision and Electronic Sensors Directorate under contract number W909MY-09-C-0001 for their generous support. \end{acknowledgments}

\appendix

\section{\label{app:hyperfine}Hyperfine}

\begin{figure}[t]
\vspace{1em}\includegraphics[width=3.5in]{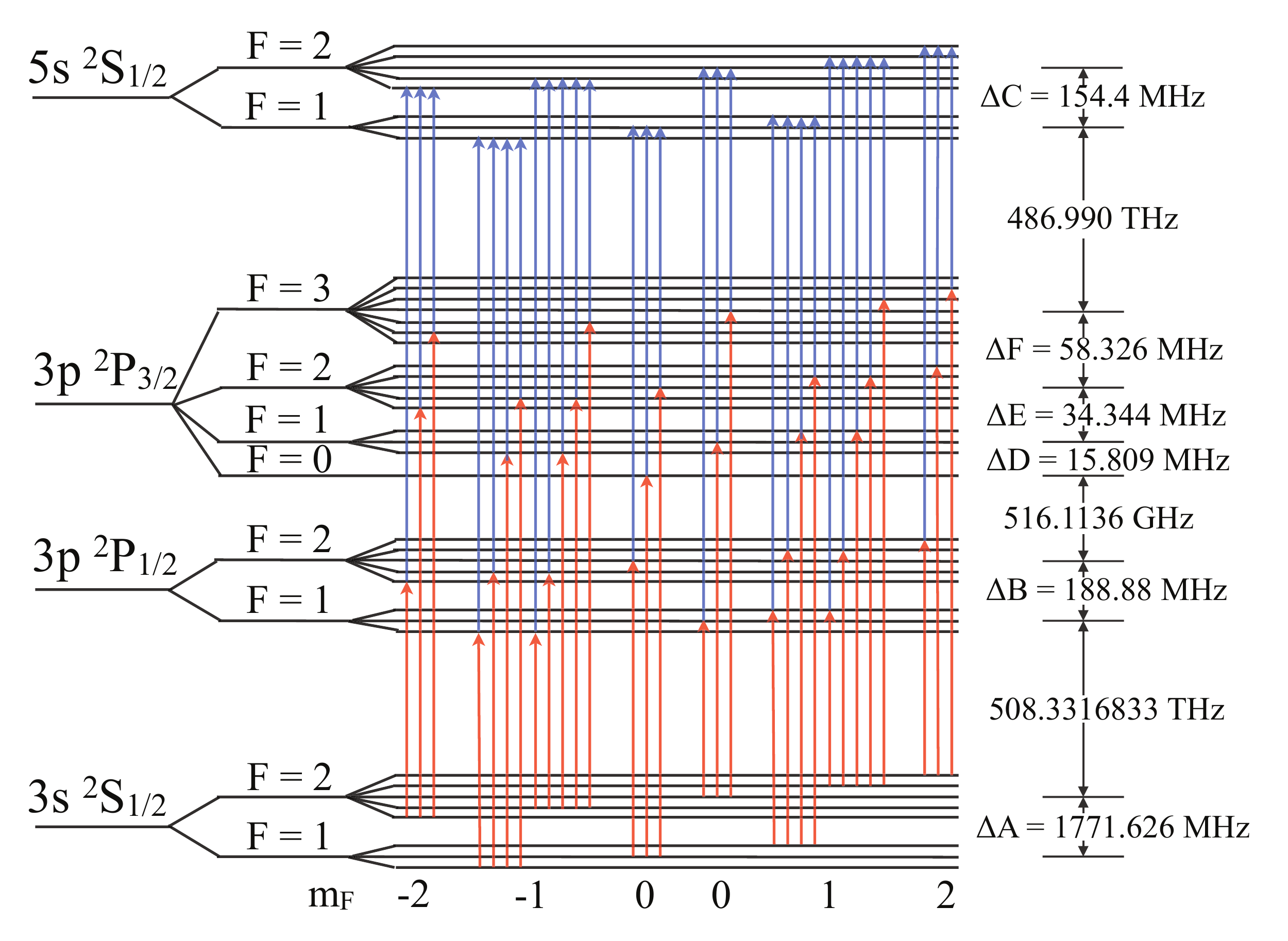}
\caption{\label{fig:energy_levels} (Color online) The hyperfine energy-levels of the 3s, $3{\rm p}\,^2{\rm P}_{1/2}$, $3{\rm p}\,^2{\rm P}_{3/2}$, and 5s states. The magnetic hyperfine sublevels are shown separated for illustration only. No external magnetic field was applied.}
\end{figure}

The hyperfine levels within the 3$\,^2{\rm S}_{1/2}$, 3$\,^2{\rm P}_{1/2}$, $3\,^2{\rm P}_{3/2}$, and $5^2S_{1/2}$ states are shown in Fig.~\ref{fig:energy_levels}. Since linearly-polarized light is used, selection rules dictate that $\Delta m_F = 0$ and $\Delta F = 0, \pm1$. Additionally, when $m_F = 0$ then $\Delta F \neq 0$. It follows that the transitions in sodium separate into six independent sets of states (one each for $m_F = \pm 2, \pm 1$, and two independent processes for $m_F = 0$) that do not interconvert during the excitation process. The six groupings of levels are shown in Fig.~\ref{fig:energy_levels}.

The Hamiltonians for the six sets of states presented here were calculated using the rotating-wave approximation. The Hamiltonians for the $m_F = \pm 2$ are shown in Eq.~\ref{eq:pm2} where the five states are labeled in ascending energy order ($\Ket{1} = \Ket{3s,F=2,m_F=\pm 2}$, $\Ket{2} = \Ket{3p \;^2{\rm P}_{1/2},2,\pm 2}$, $\Ket{3} = \Ket{3p \;^2{\rm P}_{3/2},2,\pm 2}$, $\Ket{4} = \Ket{3p \;^2{\rm P}_{3/2},3,\pm 2}$, and $\Ket{5} = \Ket{5s,2,\pm 2}$), $\Delta p$ and $\Delta s$ are the detuning of the pump and Stokes lasers from resonance, $\Delta SO$ = 516.2080 GHz, $\Omega_{ij}$ is the Rabi frequency between states $i$ and $j$, and $\Delta A$, $\Delta B$, $\Delta C$, $\Delta D$, $\Delta E$, and $\Delta F$ are given in Fig.~\ref{fig:energy_levels}. The matrix forms of the Hamiltonians for the $m_F = \pm 2$ sets of states are identical although the signs of some of the Rabi frequencies differ according the signs of the transition moments as given by Steck \cite{Steck2010}.

The $m_F = \pm 1$ sets of states are given in Eq.~\ref{eq:pm1}. As with the Hamiltonians for the $m_F = \pm 2$ states, the matrix forms of the Hamiltonians for the $m_F = \pm 1$ states are identical to each other although the signs of some of the Rabi frequencies again differ according to the transition moments given by Steck \cite{Steck2010}.

The Hamiltonians for the states with $m_F = 0$ separate into two independent sets of states as shown in Fig.~\ref{fig:energy_levels}. The Hamiltonian for the states that start in the 3s $F = 1$ state are given in Eq.~\ref{eq:01}. The Hamiltonian for the states that begin in the 3s $F = 2$ state are given in Eq.~\ref{eq:02}.

\begin{widetext}
\begin{equation}
\label{eq:pm2}
\begin{bmatrix}
-2\Delta p - \Delta A & \Omega_{12} & \Omega_{13} & \Omega_{14} & 0 \\
\Omega_{21} & - \Delta B & 0 & 0 & \Omega_{25} \\
\Omega_{31} & 0 & -2(\Delta SO + \Delta D + \Delta E) & 0 & \Omega_{35} \\
\Omega_{41} & 0 & 0 & -2(\Delta SO + \Delta D + \Delta E + \Delta F) & \Omega_{45} \\
0 & \Omega_{52} & \Omega_{53} & \Omega_{54} & 2 \Delta s - \Delta C \\
\end{bmatrix}
\end{equation}

\begin{equation}
\label{eq:pm1}
\begin{bmatrix}
\!-2\Delta p\!+\!\Delta A&0&\Omega_{13}&\Omega_{14}&\Omega_{15}&\Omega_{16}&0&0&0 \\
0&\!\!\!\!\!\!\!\!\!\!\!-2\Delta p\!-\!\Delta A\!&\Omega_{23}&\Omega_{24}&\Omega_{25}&\Omega_{26}&\Omega_{27}&0&0 \\
\Omega_{31}&\Omega_{32}&\Delta B&0&0&0&0&\Omega_{38}&\Omega_{39} \\
\Omega_{41}&\Omega_{42}&0&\!-\Delta B&0&0&0&\Omega_{48}&\Omega_{49} \\
\Omega_{51}&\Omega_{52}&0&0&\!\!\!\!-2(\Delta SO\!+\!\Delta D)&0&0&\Omega_{58}&\Omega_{59} \\
\Omega_{61}&\Omega_{62}&0&0&0&\!\!\!\!\!\!\!\!\!\!\!\!\!\!\!\!\!-2(\Delta SO\!+\!\Delta D\!+\!\Delta E)&0&\Omega_{68}&\Omega_{69} \\
0&\Omega_{72}&0&0&0&0&\!\!\!\!\!\!\!\!\!\!\!\!\!\!\!\!-2(\Delta SO\!+\!\Delta D\!+\!\Delta E\!+\!\Delta F)&0&\Omega_{79} \\
0&0&\Omega_{83}&\Omega_{84}&\Omega_{85}&\Omega_{86}&0&\!\!\!\!\!\!2\Delta s\!+\!\Delta C&0\\
0&0&\Omega_{93}&\Omega_{94}&\Omega_{95}&\Omega_{96}&\Omega_{97}&0&\!\!\!\!\!\!2\Delta s\!-\! \Delta C \\
\end{bmatrix}
\end{equation}

\begin{equation}
\label{eq:01}
\begin{bmatrix}
-2\Delta p + \Delta A & \Omega_{12} & \Omega_{13} & \Omega_{14} & 0 \\
\Omega_{21} & - \Delta B & 0 & 0 & \Omega_{25} \\
\Omega_{31} & 0 & -2\Delta SO & 0 & \Omega_{35} \\
\Omega_{41} & 0 & 0 & -2(\Delta SO + \Delta D + \Delta E) & \Omega_{45} \\
0 & \Omega_{52} & \Omega_{53} & \Omega_{54} & 2 \Delta s + \Delta C \\
\end{bmatrix}
\end{equation}

\begin{equation}
\label{eq:02}
\begin{bmatrix}
-2\Delta p - \Delta A & \Omega_{12} & \Omega_{13} & \Omega_{14} & 0 \\
\Omega_{21} & + \Delta B & 0 & 0 & \Omega_{25} \\
\Omega_{31} & 0 & -2(\Delta SO + \Delta D) & 0 & \Omega_{35} \\
\Omega_{41} & 0 & 0 & -2(\Delta SO + \Delta D + \Delta E + \Delta F) & \Omega_{45} \\
0 & \Omega_{52} & \Omega_{53} & \Omega_{54} &  2 \Delta s - \Delta C \\
\end{bmatrix}
\end{equation}

\end{widetext}

\section{\label{sec:modelfluor}Modeling detected fluorescence}

\subsection{\label{sec:modelfluorparam} Determination of Rabi frequencies}

Inasmuch as our laser pulses are Gaussian in space and time, the pulse intensity as a function of the peak electric field strength, time, and radial distance from the pulse center is given by:

\begin{align}
\label{eq:Ioft}
I(E_0\, , t\, ,r) &= (1/2)\epsilon_0 c E_0^2 e^{-2\left(t/\Delta \tau\right)^2}e^{-2\left(r/W\right)^2}
\end{align}

\noindent where $\epsilon_0$ is the permittivity of free space, $c$ is the speed of light in vacuum, $E_0$ is the peak electric field magnitude in both position and time, $\Delta \tau$ is the measured $1/e^2$ width of the pulse duration, and $W$ is beam radius at $1/e^2$ of its maximum. The temporal full width at half maximum is given by $FWHM=\sqrt{2\, ln(2)}\Delta \tau$. Integrating the intensity, $I(E_0\, , t\, , r)$, over both time and area provides the energy as a function of $E_0$, $W$, and $\Delta \tau$:

\begin{align}
\begin{split}
\label{eq:Iint}
U(E_0)&=\int_{-\infty}^{\infty} dt \int_0^{2 \pi}d\phi \int_0^{\infty} dr \, r \,\,I(E_0\, , t\, ,r) \\
&=(\pi/2)^{3/2}
\epsilon_0 c \Delta \tau W^2 E_0^2 \,.
\end{split}
\end{align}

\noindent Eq.~\ref{eq:Iint} is solved for $E_0$ and inserted into the expression for the maximum Rabi frequency:

\begin{align}
\Omega_0 &= \frac{\mu \cdot E_0}{\hbar} = \frac{\mu\cdot \sqrt{
\frac{
2^{3/2}U(E_0)}{\pi^{3/2} \epsilon_0 c \Delta \tau W^2}}}{\hbar}\label{eq:omeganot}
\end{align}

\noindent where $\mu$ is the transition dipole matrix element for the transition of interest. The parameters $U(E_0)$, $\Delta \tau$ and $W$ are measured experimentally and $\Delta \tau$ and $W$ are listed in Table~\ref{tab:pulse_parameters}. We now integrate the Rabi frequency over time, using the envelope width of the electric field, as opposed to the intensity as in the integral in Eq.~\ref{eq:Iint}:

\begin{align}
\Omega_{int} &= \int_{-\infty}^{\infty}\,\Omega_0 \cdot e^{-\left(t/\Delta \tau \right)^2}\,dt \  = \  \sqrt{\pi}\, \Delta \tau\, \Omega_0\label{eq:int_rabi_freq}
\end{align} 

\noindent which gives us the peak integrated Rabi frequency at the center of the Gaussian beam. The integrated Rabi frequencies are then calculated over the entire spatial Gaussian envelopes of both the pump and Stokes beams, which is the integrated Rabi frequency times the spatial dependence of the beam given in Eq.~\ref{eq:Ioft}:

\begin{align}
\Omega_{i}(r) &= \  \sqrt{\pi}\, \Delta \tau\, \Omega_0 \,e^{-2\left(r/W \right)^2}.\label{eq:int_rabi_freq_r}
\end{align} 

\noindent When $\Omega_{i}$ in Eq.~\ref{eq:int_rabi_freq_r} refers specifically to the pump or Stokes beams a subscript p or s is added in equations that follow. Eq.~\ref{eq:int_rabi_freq_r} is used to calculate transfer efficiencies (STIRAP, SEP, Autler-Townes, ...) as a function of position within the beams. Finally, the amount of light that enters the spectrograph as a function of position in the beam is calculated and spatially integrated over the beam to yield the quantity of light collected by the spectrograph. The details of this process are outlined in Appendix~\ref{app:Eff}.

For purposes of calibration, we investigate how strongly the falloff of intensity in the spatial beam profile for both the pump and Stokes beams contributes to the STIRAP efficiency measured in sodium vapor. It is common to use the time-integrated Rabi frequency, or pulse area in Eq.~\ref{eq:int_rabi_freq} as a measure of the strength of the beams. It is important to note that this measure only refers to the peak of the spatial Gaussian beam and does not take into account the wings of the Gaussian and the related lower intensities as the profile falls off towards zero.

\begin{figure}[t]
\includegraphics[width=3.5in]{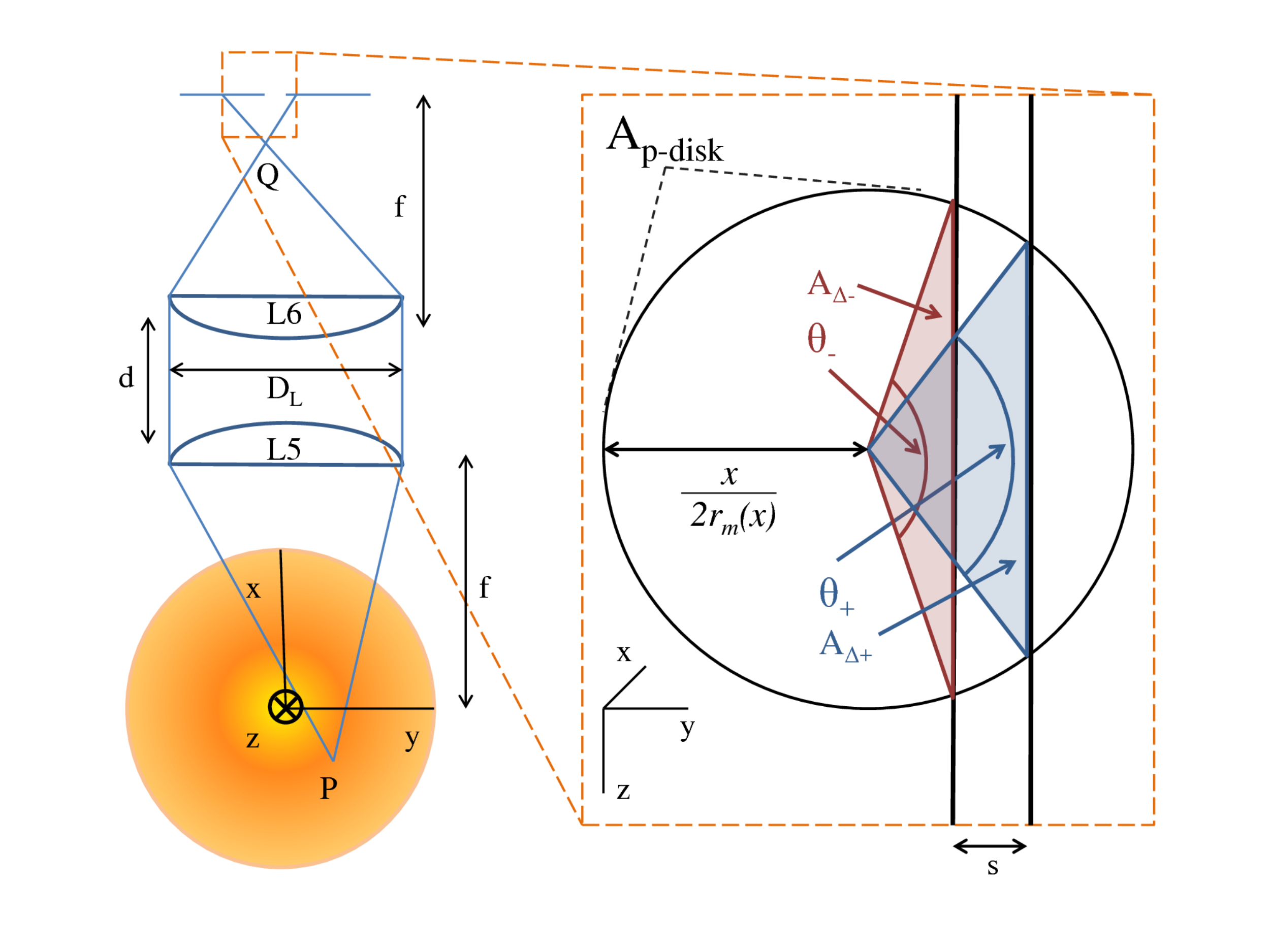}
\caption{\label{fig:Pxy} (Color online) Collection of fluorescence into a spectrograph. Pump and Stokes beams travel into the paper at the lower-left corner giving rise to sodium fluorescence. Fluorescence is collected by two lenses and focused on the entrance slit of the spectrograph. Since the fluorescence from various locations within the beam cannot all be brought to a focus on the entrance slit, the center of the beam is imaged perfectly on the center of the slit. Fluorescence from the representative point P behind and to the right of beam center forms an unfocused circle of light to the left of the entrance slit as shown in the expanded view to the right. Experimental parameters from the diagram are as follows: spectrograph slit width $s$ = 0.028 mm, lens focal length f = 203.2 mm, lens diameter $D_{L}$ = 50.8 mm, lens separation $d=1-2$ cm.
}
\end{figure}

\subsection{\label{sec:Pxy}Probability of photon detection - P$(x,y)$}

To determine how the spatial beam profile contributes to the measured fluorescence, it is first necessary to model the area of fluorescence generated by the overlapping beams in the sodium vapor and the collection of light along the optical path to the spectrograph entrance slit. By doing so we can predict the probability that a spontaneously emitted fluorescent photon will enter the spectrograph as a function of its emission position in the beam profile. In Fig.~\ref{fig:Pxy}, fluorescence from the center of the beam is imaged onto the center of the slit; fluorescence emitted from the point P behind and to the right of the center of the beam is brought to a focus in front and to the left of the entrance slit. Since the light from P is brought to a focus in front of the slit, it then diverges, forming a circle by the time it reaches the plane of the slit. Note that fluorescence emerging from points in front of the center of the beam would also form a circle at the plane of the slit since they would be be intercepted while converging to a point behind the slit. The circle of light on the entrance slit is shown in more detail as an expanded view on the right side of Fig.~\ref{fig:Pxy}. The fraction of this circle that is within the slit enters the spectrograph. The analysis that follows illustrates how this fraction is calculated.

Two assumptions are inherent in determining the probability of photon detection:

\begin{enumerate}
\item Emitted fluorescent photons are assumed to be radially symmetric about the z-axis (see Fig.~\ref{fig:Pxy}.)
\item Emission is independent of distance $z$ along the beams. Moreover the length of the slit is much greater than its width so that effects due to the ends of the slit are small. In addition, the diameters of the lenses are much greater than the length of the slit.
\end{enumerate}


The optical imaging of photons onto the slit along the x-axis through a pair of thin lenses is given by:

\begin{equation}
\label{eq:thinlens}
q(p)=\frac{df(p-f)-pf^{2}}{(d-f)(p-f)-pf}
\end{equation}

\noindent where $q(p)$ is the image distance as a function of the object distance, $p$, for a point source at position $P$ and image position at $Q$, $f$ is the focal length of the plano-convex lenses L4 and L5, and $d$ is the distance of separation between the two lenses. The center of the beam profile and the center of the spectrograph slit are located at the object and image focal lengths, respectively. See Fig.~\ref{fig:Pxy} for a visualization of the optical path.

From Eq.~\ref{eq:thinlens}, we find that if $p\approx f$, the x-position of the final image is near the entrance slit of the spectrograph. Setting $p=f-x$ in Eq.~\ref{eq:thinlens}, the Maclaurin series expansion in $x$ is 

\begin{equation}
\label{eq:thinlenslinear}
q_{a}(p)= f+x+\left(\frac{2}{f}-\frac{d}{f^2}\right)x^2+ O(x^3).
\end{equation}

\noindent where $q_{a}(p)$ is the approximate image position along the x-axis. The signs of the terms in Eq.~\ref{eq:thinlens} reflect the object and image x-axis directions in Fig.~\ref{fig:Pxy}. It is important to note that the p-values used in this calculation were between $f\pm 3$mm, which makes the first two terms of Eq.~\ref{eq:thinlens} accurate to $2.7\%$. This error corresponds to distances of $\pm 7.7\,\sigma$ for the pump pulse and $\pm 8.6\, \sigma$ for the Stokes pulse where $\sigma$ is the standard deviation with respect to the measured Gaussian spatial pulse.

Using the constant and linear terms of Eq.~\ref{eq:thinlens}, we observe a direct mapping along the x-axis for source photons in the beam and image photons at the slit, i.e. a photon emitted 1 mm away from beam center, in a negative direction along the optical axis will be imaged 1 mm in front of the slit. Mapping the y-axis is simple: $y_{image} = -(q/p)y_{source}\approx -y_{source}$.

If one takes the beam profile to be a continuous collection of point sources, it falls next to determine how an emission from each position in the Gaussian beam profile, $(x,y)$, maps to a disk projected onto the spectrograph slit, and the fraction of the disk that enters the spectrograph. 

In order to simplify calculations, we define

\begin{equation}
\label{eq:modfnumber}
r_m(x) = \frac{f-x}{D_{L}}
\end{equation}

\noindent where $D_{L}$ is the diameter of the lenses. As $x\rightarrow 0$, the image is formed at the focal point of the lens and Eq.~\ref{eq:modfnumber} reduces to the f-number of the lens.

\begin{figure}[t]
\includegraphics[width=3.5in]{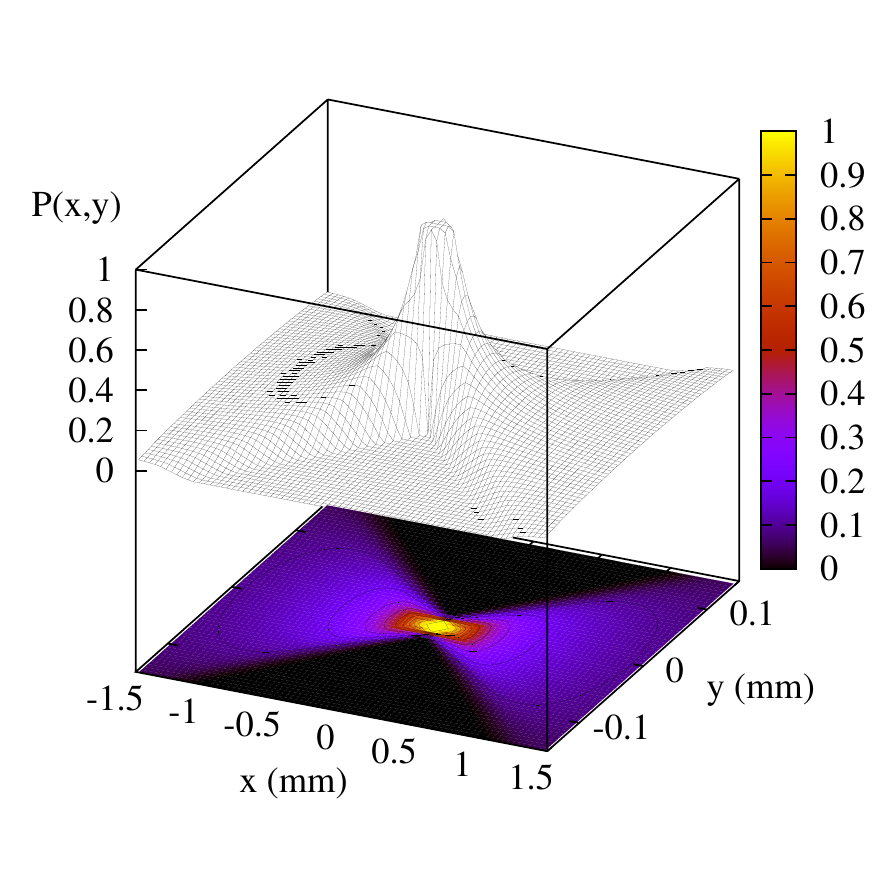}\\
\caption{\label{fig:bowtiecontour} (Color online) The bow-tie region defined in Eq.~\ref{eq:bowtie}. The colorbar and $P(x,y)$ represent the probability of a photon emitted at point $(x,y)$ to enter spectrograph slit assuming pump and Stokes aligned at the origin along z-axis. Note the scale is an order of magnitude larger on the x-axis than y-axis.}
\end{figure}

To first order in $x$ the spatial boundary condition for a photon emitted from an area in the beam profile with a non-zero detection probability is given by:

\begin{equation}
\label{eq:bowtie}
|y| \leq \frac{x+r_m(x)\cdot s}{2\,r_m(x)}
\end{equation} 

\noindent where $x$ and $y$ correspond to the axes in Fig.~\ref{fig:Pxy} and $s$ is the spectrograph slit width. The boundary condition seen in Eq.~\ref{eq:bowtie} is shown in Fig.~\ref{fig:bowtiecontour} as the boundary making up the bow-tie shape. 

The probability $\operatorname{P}(x,y)$ that a photon emitted from $(x,y)$ in the beam profile (and collected by the lenses) will enter the spectrograph is equal to the area of the projected disk that overlaps the slit divided by the total disk area (See Fig.~\ref{fig:Pxy}.) This is calculated as follows.

\begin{align}
\label{eq:Pxy}
\operatorname{P}(x,y) &= \frac{A_{\text{near}}(x,y) - A_{\text{far}}(x,y)}{A_{\text{p-disk}}(x)}
\end{align}
\noindent where
\begin{align}
A_{\text{p-disk}}(x) &= \pi\cdot \left(\frac{x}{2\cdot r_m(x)}\right)^2\\
A_{\text{near}}(x,y) &= \frac{\theta_{-}(x,y)}{2\pi} \cdot A_{\text{p-disk}}(x) - A_{\Delta -}(x,y)\\
A_{\text{far}}(x,y) &= \frac{\theta_{+}(x,y)}{2\pi}\cdot A_{\text{p-disk}}(x) - A_{\Delta +}(x,y)
\end{align}

\noindent with $\theta_{\pm}(x,y)$ and $A_{\pm}(x,y)$ given by:

\begin{align}
\theta_{\pm}(x,y) &= 2\cdot \cos^{-1}\left[\frac{2\cdot r_m(x)\cdot(y \pm \frac{s}{2})}{|x|}\right ]\\
A_{\Delta\pm}(x,y) &= \left(y \pm \frac{s}{2}\right)\cdot \sqrt{\left(\frac{A_{\text{p-disk}}(x)}{\pi}\right ) - \left(y \pm \frac{s}{2}\right )^2}.
\end{align}

\noindent The quantities $A_{\text{p-disk}}(x)$, $A_{\Delta \pm}(x,y)$ and $\theta_{\pm}(x,y)$ are shown in Fig.~\ref{fig:Pxy}. $A_{\text{near}}(x,y)$ and $A_{\text{far}}(x,y)$ are respectively the circular segments to the right of the left and right edges of the slit. These quantities are calculated by subtracting the triangular areas, $A_{\Delta -}$ and $A_{\Delta +}$ respectively, from the circular sectors formed by the angles $\theta_{-}$ and $\theta_{+}$ respectively. Due to symmetry, the calculations need only be performed for one quadrant in the xy-plane which was arbitrarily chosen as the second ($x\leq 0, y \geq 0$). Furthermore, it is important to note that the following condition is necessary for non-zero values of $\theta_{\pm}$ and $A_{\Delta \pm}$:

\begin{align}
0 &\leq y < \frac{|x|}{2\cdot r_m(x)}\mp\frac{s}{2}
\end{align}

\noindent where the zero on the left represents our choice of the second quadrant. The probability $\operatorname{P}(x,y)$ is shown in Fig.~\ref{fig:bowtiecontour}.

\begin{figure}[h]
\includegraphics[width=3.5in]{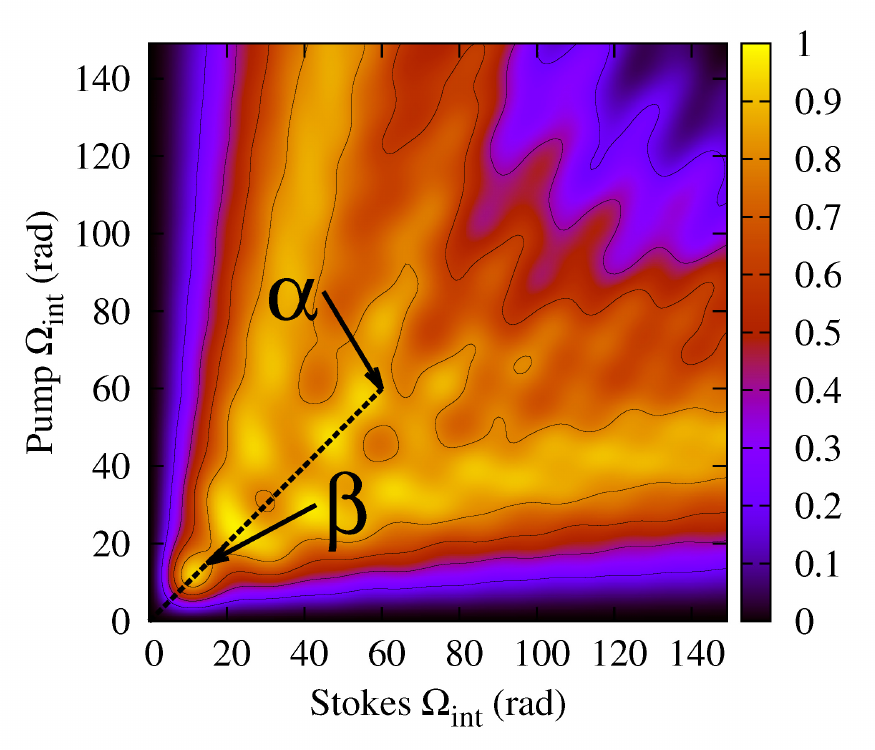}\\
\vspace{-4em}
\includegraphics[width=3.5in]{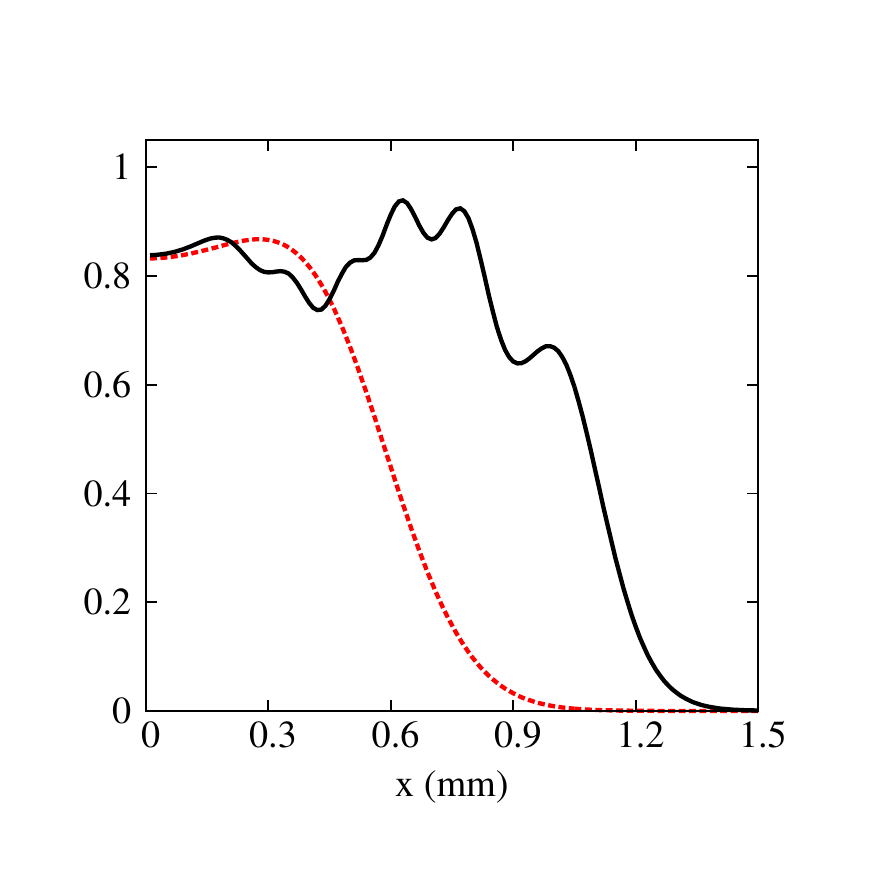}
\vspace{-3.5em}
\caption{\label{fig:d1_cont_line}(Color online) {\bf Top}: Transfer efficiency from the 3s to the 5s state with the lasers tuned through the D1 transition as a function of Stokes and pump integrated Rabi frequencies. The arrows mark two example maximum Rabi frequencies, $\alpha$ and $\beta$. The black dashed line illustrates the decreasing intensity (toward the origin) while moving from the center toward the wings of the Gaussian beam envelopes for equal Stokes and pump spatial Gaussian beam widths. {\bf Bottom}: Radial plots of Eff$(\Omega_s,\Omega_p)$ for $\alpha$ and $\beta$, marked in the upper panel. Eff$_{\alpha}$ is represented by the solid black line and Eff$_{\beta}$ is represented by the dashed red line.} 
\end{figure}

Note that for our experimental setup, the center of the beam profile, and a small diamond-shaped area around it, shows probability of a photon entering the spectrograph as 1.0; this probability initially falls off more rapidly than $x^{-2}$ along the x-direction, but at relatively large distances the behavior asymptotically approaches $x^{-1}$. It is interesting to note that vertical slices of equal width in the probability given in Fig.~\ref{fig:bowtiecontour} have equal areas. This indicates that even though the intense center of the beam is imaged onto the spectrograph entrance slit, the probability of collecting photons at any location $x$ (integrating over $y$) in front of or behind the center of the beam is equally likely, resulting in the collection of many photons in the wings of the beam where the intensity is much smaller.

\subsection{\label{app:Eff}Combining transfer efficiency}


Numerically solving the Hamiltonian given in Eq.~\ref{eq:fine} or the set of Hamiltonians in Appendix~\ref{app:hyperfine} allows us to map out the transfer efficiency to the 5s state as a function of the pump and Stokes integrated Rabi frequencies, as shown in Fig.~\ref{fig:Energy_scan_2}. The transfer efficiency as a function of the integrated Rabi frequencies allows us to calculate the transfer efficiency as a function of position $(x,y)$ within the beams $\operatorname{Eff} [\Omega_{is}(x,y),\Omega_{ip}(x,y)]$.

Integrating the product of the transfer efficiency for a given set of Rabi frequencies, Eff, and probability of detection, $P(x,y)$, over the spatial dimensions of the Gaussian pulse provides a dimensionless measure, $\epsilon(\Omega_{is},\Omega_{ip})$, of the expected level of fluorescence detection at a given set of pump and Stokes pulse energies:

\begin{equation}
\label{eq:epsilon}
\epsilon(\Omega_{is}, \Omega_{ip}) = \int \text{Eff}[\Omega_{is}(x,y),\Omega_{ip}(x,y)]\cdot\text{P(x,y)}\,dx\,dy
\end{equation}

The upper contour plot of Fig.~\ref{fig:d1_cont_line} displays a reduced region of the top panel of Fig.~\ref{fig:Energy_scan_2} (the STIRAP transfer efficiency through  $3\,^2\operatorname{P}_{1/2}$). The two arrows indicate maximum integrated Rabi frequencies at the center of laser pulses for $(\Omega_{is},\Omega_{ip}) \approx (60.0, 60.0)$ rad at $\alpha$ and $(\Omega_{is},\Omega_{ip})\approx (15.0, 15.0)$ rad at $\beta$. The black line that extends from $\alpha$ or $\beta$ to the origin illustrates the path in Rabi-frequency space obtained from Eq.~\ref{eq:int_rabi_freq_r} as the radial distance from the center of the pulses, $r$, increases. The bottom panel of Fig.~\ref{fig:d1_cont_line} displays $\operatorname{Eff}(\O mega_{is},\Omega_{ip})$ as a function of $r$ for $\alpha$ and $\beta$ in the top panel. 

It is interesting to note that there are intensities along these pulse-pair envelopes that contribute more strongly than the pulse peak does to a transition, i.e. higher Eff$(\Omega_{is}, \Omega_{ip})$ values. The higher intensity pair of pulses, $\alpha$, shows more complexity along the Gaussian envelope than the lower intensity pair, $\beta$, as the former follows a longer path along the black line in the upper contour plot due to the less rapid reduction in intensity from the Gaussian wings. It therefore passes over more of the contoured structure in the upper mapping. 

The quantities plotted in the bottom panel of Fig.~\ref{fig:d1_cont_line} show clearly that even after combining the rapid radial reduction of photon detection probability from Sec.~\ref{sec:Pxy}, the photons emitted from the wings of the Gaussian areas of fluorescence can still contribute strongly to the detected fluorescence.


\bibliographystyle{unsrt}

\bibliographystyle{unsrt}


\end{document}